\documentclass[fleqn,usenatbib]{mnras}

\usepackage{newtxtext,newtxmath}

\usepackage[T1]{fontenc}
\usepackage{ae,aecompl}

\usepackage{graphicx}	
\usepackage{amsmath}	
\usepackage{amssymb}	

\title[Simulating AM 2229-735 and its polar structure]{Simulating the
  galactic system in interaction AM 2229-735 and the formation of its
  polar structure}

\author[Luis F. Quiroga et al.]{
Luis F. Quiroga,$^{1}$\thanks{E-mail: luis.quiroga@udea.edu.co}
J. C. Mu\~noz-Cuartas,$^{1}$
I. Rodrigues$^{2}$
and Noam I. Libeskind$^{3,4}$
\\
$^{1}$Instituto de F\'isica, Universidad de Antioquia; Calle. 67
No. 53 - 108, A. A. 1226; Medell\'in, Colombia\\
$^{2}$Universidade do Vale do Para\'iba; Av. Shishima Hifumi, 2911,
CEP:12.244-000; S\~ao Jos\'e dos Campos, Brazil\\
$^{3}$Leibniz-Institut f\"ur Astrophysik Potsdam; An der Sternwarte 16,
14482; Potsdam, Germany\\
$^{4}$l'Institut de Physique Nucl\'eaire de Lyon (IPNL), University of
Lyon; UCB Lyon 1/CNRS/IN2P3; Lyon, France}

\date{Accepted 2019 October 11. Received 2019 October 11; in original form 2019 July 27}

\pubyear{2019}

\begin{document}
\label{firstpage}
\pagerange{\pageref{firstpage}--\pageref{lastpage}}
\maketitle

\begin{abstract}
We study the formation of polar ring galaxies via minor mergers. We
used N-body+hydrodynamics simulations to reproduce the dynamics of the
observed system AM 2229-735 that is a minor merger whose interaction
signals are those of a progenitor for a polar ring galaxy. We used the
observational information of the system to get initial conditions for
the orbit and numerical realisations of the galaxies to run the
simulations. Our simulations reproduce the global characteristics of
interaction observed in the system such as arms and a material bridge
connecting the galaxies. As a merger remnant, we found a quasi-stable
and self gravitating planar tidal stream with dark matter, stars and
gas orbiting in a plane approximately perpendicular to the main galaxy
disk leading in the future to a polar ring galaxy. We studied the
dynamical conditions of the polar structure and found evidence
suggesting that this kind of merger remnant can settle down in a
disk-like structure with isothermal support, providing inspiring
evidence about the process of formation of galactic disks and
providing a potentially independent scenario to study the presence of
dark matter in this kind of galaxies.
\end{abstract}




\section{Introduction}

Any successful cosmological model should be able to account not only
for the observed matter distribution in the Universe but also for its
observed dynamics. In the $\Lambda$CDM scenario, the currently
observed structures of matter are the result of a hierarchical process
of structure formation, where primordial over-densities accreted dark
matter via gravitational instability to form even bigger matter
overdensities called dark matter halos. The baryonic matter is
captured by these over densities falling in the gravitational
potential well of halos \citep{Benson2010}. In this process it is
supposed that angular momentum is conserved, thereby gas clouds get
angular momentum from its orbital angular momentum obtaining a net
rotation; additionally, while gas clouds collapse they cool and form
stars. In this case the final result is a dark matter halo with an
axis-symmetric gaseous and stellar structure in its centre called
galactic disk. There are galaxies where mass accretion is violent and
the gas and stars form a structure with spheroidal symmetry with
almost zero net rotation, they define what we call elliptical galaxies
\citep{Mo2010, Benson2010}.

In this cosmological context, galaxies undergo constant collisions
leading to their evolution and to the great variety of galaxies
observed at different epochs of the Universe. This hierarchical model
of structure formation has been very successful to reproduce the
distribution of matter in the observed universe with its main
characteristics. However, still there are several open problems, one
of these are the formation of disk galaxies with a complex
substructure such as rings, arms, and polar structures.

The vast variety of morphologies and matter distribution product of
gravitational interaction in galaxies is evident in observations
\citep{Arp1977, Arp1987, Martinez-delgado2010, Morishita2014,
  Shibuya2015, Shibuya2016}. One can find elliptical and S0 galaxies
with inclined disks and rings, galaxies with streams of infalling
material accreted from orbiting satellites, galaxies with shells,
ripples and perturbed disks, etc.

A very interesting example of a galactic system are the polar ring
galaxies (PRGs). They are exotic systems that are believed to be the
result of a galactic interaction in a privileged direction. They are
mainly characterised by a central S0 galaxy almost perpendicularly
surrounded by an accreted structure of gas, dust, and stars settled in
a quasi-equilibrium configuration (the galaxy NGC 4650 is an example
of PRG, see \cite{Whitmore1990, Reshetnikov1997, Laurikainen2011,
  Laurikainen2013}). Since an important fraction ($\sim$20-25$\%$) of
S0 galaxies show some polar structure like disks, rings or incomplete
versions of them, to understand how this systems formed will extend
our knowledge on the formation and evolution of
galaxies. Particularly, they could be key scenarios to understand how
galactic interactions happens, how mergers affect galactic morphology
and kinematics, to understand the shape of gravitational potential in
galaxies, the shape of dark matter halos, induced variations in star
formation rate and metal distributions, among many other open
questions on the evolution of galaxies.

Several models have been proposed to explain the formation and
structure of polar ring galaxies. Accretion models propose that the
material in the ring comes from interactions with a gas rich galaxy
forming a ring around the host galaxy \citep{Schweizer1983,
  Whitmore1990, Steiman-Cameron1988, Steiman-Cameron1990}. In merging
models  all material is stripped from a secondary (minor) galaxy
forming a ring perpendicular to the major semi-axis of the host galaxy
\citep{Bekki1998, Bournaud2003}. Other models propose that smoothly
infalling material forms a ring while accreted from cosmic filaments
\citep{Maccio2006, Brook2008}. All these models have been tested with
numerical simulations that have succeeded to reproduce many of the
properties observed in PRGs, these results together with results of
observations have found correlations between the properties of the
host galaxy with those of its polar structure \citep{Reshetnikov1997,
  Bekki1998}.
  
Currently, there are still open questions concerning the formation of
PRGs. To mention some of them, we do not know yet which would be the
expected frequency of PRGs, how is the kinematics of stars forming in
these structures, It is not yet clear what is the origin of the
gradients of metallicity observed in PRGs, what are the formation
mechanism of massive self-gravitating stellar-gaseous polar rings,
among many others \citep{Moiseev2011,Combes2013}. In that sense, the
use of hydrodynamical simulations to reproduce observed interacting
systems candidate to form polar structures becomes interesting
astrophysical laboratories. These simulations, tuned to reproduce a
specific observed system, would provide useful information to
understand the influence of the morphological properties of the
galaxies and the merger orbit in the final properties of the polar
ring or disk formed. One of these candidates, is the system AM
2229-735 \citep{FreitasLemes2014} that is described below and will be
the object of study in this work.

In this work we want to study the problem of the formation of polar
ring galaxies in a merging model where a satellite galaxy is stripped
by a host galaxy and the produced tidal stream is settled forming a
polar structure around this host. For that we study the evolution of
the real system AM 2229-735 using its observational information to
setup N-body+hydrodynamical simulations with radiative cooling, star
formation and supernovae feedback. In these simulations we found a
galaxy with a polar structure around it and we studied with detail the
properties and evolution of tidal stream forming the structure.

This paper is outlined as follows. In Section \ref{sec:observations}
we present the state of the art of observations of the system AM
2229-735 and the observational data that we use for this work. The
method to get initial conditions for simulations from observations in
Section \ref{sec:observationstosimulations}. The numerical details of
simulations are presented in Section \ref{sec:simulations} and an
analysis of polar structure formed in Section
\ref{sec:results}. Finally, we present our conclusions in Section
\ref{sec:conclusions}.

\section{Observations}
\label{sec:observations}

From its first report in ``A CATALOGUE OF SOUTHERN PECULIAR GALAXIES
AND ASSOCIATIONS'' \citep{Arp1977, Arp1987}, the system AM 2229-735
have been studied by several authors, this has allowed to collect many
data that will be useful to configure initial conditions for the
simulations and to compare their results with observations to get
clues about its evolution and fate.

\cite{Ferreiro2004} used images from CTIO 0.9m telescope plus spectra
from \cite{Donzelli1997} to study the effects of the interaction on
the integrated photometric properties and star formation activity,
they determined that AM 2229-735 is conformed by a main disk galaxy
and a compact elliptical companion galaxy. The main galaxy, with an
exponential luminosity profile, is very perturbed and exhibits several
regions of active star formation. The galaxies are connected with a
luminous bridge that contains 17\% of the total luminosity of the
system. Later, \cite{Ferreiro2008} studied the properties of those HII
regions. A comparison of these with other HII regions in normal and
isolated galaxies and tidal  dwarf  galaxies candidates was
done. They estimated the age, star formation rate and nuclear
mass of both galaxies and of six HII regions of the main
galaxy. They found that in the satellite galaxy there is H$\alpha$
emission only in its nucleus and that in the tip of the bridge that
connects the galaxies there is a special region with the youngest
stellar population of the system.

\begin{figure}
   \centering
   \includegraphics[scale=0.76,angle=270]{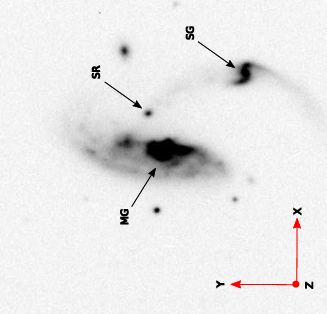}
   \caption{GMOS-S \textit{r}-band image for system AM 2229-735. North
     is up, East is to the right with, the image size of $58.4''\times
     58.4''$ corresponding to a scale of
     $0.146\:''/pix=0.169\:kpc/pix$.}
 \label{fig:AM}
 \end{figure}

\cite{FreitasLemes2014} made a detailed study of AM 2229-735 using
acquisition images and spectroscopy data obtained at Gemini South (Chile) and
Pico dos Dias (Brazil) telescopes. They analysed broad-band images and 
long-slit spectroscopy. These images, with
better resolution than used in previous works, show clearly that the
satellite is not an spheroidal but a disk galaxy (see
Fig. \ref{fig:AM}). These observations allows the detection of a tail
and counter-tail arc-shaped features. The southern region of the
main galaxy is bluer than the northern one, its centre has a
bar. The companion galaxy, almost face-on along the line-of-sight, is
in general bluer than the host and exhibits an U-shaped radial velocity
profile typical of interacting galaxies.

Recently, \cite{Krabbe2017} used a sample of nine interacting galaxies
to study the relation between the gravitational interaction and the
stellar populations. In this study, the main galaxy disk of AM
2229-735 has two interesting HII regions dominated by young and
intermediate age stellar populations probably formed during the
interaction. Finally, as an important particular case for this work,
in \cite{FreitasLemes2014} AM 2229-735 is proposed as a progenitor of
a polar ring galaxy because its morphology is similar to the later
instants observed in simulations of polar ring galaxies shown in
\cite{Bournaud2003,Reshetnikov2006}.

To reproduce the interaction of AM 2229-735 it is necessary to obtain,
from observations, information about the galaxies in the pair. For
that, we used public acquisition images in \textit{r}-band
superimposed on Gemini Multi-Objects Spectrograph (GMOS-S) acquired as
part of the bad weather program GS-2006A-DD-6 in Gemini South,
Chile. Aims and results of these observations are detailed in the
paper series \cite{Krabbe2014, Rosa2014, Krabbe2017}. Figure
\ref{fig:AM} shows an image of the system, one can see a main disk
galaxy (hereafter MG) at its centre, a satellite galaxy (hereafter
SG), a material bridge that connects both galaxies, a special luminous
region (hereafter SR) on this bridge and other objects that we assume
are field objects.

\begin{table*}
  \centering
  \caption{Masses and scale lengths for each galactic component for
    the main (MG) and satellite (SG) galaxies of system AM 2229-735.}
  \label{tab:galaxy_properties}
  \begin{tabular}{lcccc}
    \hline
    Component & Masses MG ($10^{10}\:M_{\odot}/h$) & Scale MG ($kpc/h$)& Masses SG ($10^{10}\:M_{\odot}/h$)& Scale SG ($kpc/h$)\\
    \hline
    Halo         & 70.19 & 23.22 & 3.05 & 9.05 \\
    Gas disk     & 0.16  & 3.40  & 0.03 & 1.02 \\
    Stellar disk & 5.59  & 3.40  & 0.11 & 1.02 \\
    Bulge        & 2.39  & 0.68  & 0.03 & 0.20 \\
\hline
\end{tabular}
\end{table*}

\section{Initial conditions: From observations to simulations}
\label{sec:observationstosimulations}

In this section, the procedures to estimate basic geometrical and
dynamical properties of AM 2229-735 are described. In this part of the
work, different strategies and techniques were combined to obtain all
the parameters required to build two numerical disk galaxies with
physical properties in agreement with the observations of MG and
SG. Finally, a numerical procedure was developed to obtain an orbit
that reproduce the state of the interaction observed for the system.

\subsection{Properties of disk galaxies}
\label{sec:properties}
 
We define a coordinate system for this work, with the origin in the
centre of MG (defined as the position of the maximum surface
brightness), coordinate axes are oriented with the X-axis to the right
along the horizontal, Y-axis points up along the vertical and Z-axis
pints towards the observer as illustrated in \autoref{fig:AM}.

Surface photometry analysis was performed using \textit{ellipse} task
in IRAF. Fitting the disk isophotes to an exponential profile we got
radial disk scale lengths for both MG $3.37\:kpc/h$ and SG
$1.0\:kpc/h$. In our coordinate frame, $PA$ is defined as the angular
position of disk major semi-axis from the Y-axis being positive
counterclockwise. We found this angle as the mean of ellipses fitted
for isophotes of each disk excluding the central and outer ellipses to
avoid the contribution of galactic bulge and tidally disturbed
arms. The mean $PA$ obtained this way are $-1.74^\circ$ for MG and
$-59.57^\circ$ for SG.  We define the inclination $i$, as the angle
between each disk and the XY plane (positive counterclockwise). This
angle can be found using the mean semi-major axis $a$ and semi-minor
axis $b$ of the disk as $i=\cos^{-1} \left( b/a\right)$, where $a$ and
$b$ can be estimated from the elliptical isophotes found with
\textit{ellipse}. The mean inclination, in the case of AM 2229-735 is
$-63.55^\circ$ for MG and $41.75^\circ$ for SG.

To complete the basic information to build the numerical models for
the galaxies in the pair, it is necessary to estimate the mass of each
galactic component: stellar and gaseous disks, bulge and halo. In what
follows, we will describe how we estimate each mass component for each
galaxy.

To calculate stellar mass ($M_*$) for each galaxy we used the
mass-to-light ratio based in \cite{Bell2003}. They estimated galaxy
luminosity and stellar mass functions in the local universe using data
of a large sample of galaxies from the Two Micron All Sky Survey
(2MASS) and the Sloan Digital Sky Survey (SDSS). Then they estimated
present-day stellar mass-to-light ratios and the mean stellar mass can
be estimated as

\begin{equation}
  \log(M_{*}) = -0.306 + 1.097(g-r)^0 - \epsilon + \log(L_r),
  \label{eq:stellar_mass}
\end{equation}

\noindent here, the luminosity in $r$-band ($L_r$) is required to
estimate $M*$ and $(g-r)^0$ is the galaxy colour corrected to
$z=0$. Because the filters used in the images of AM 2229-735 are
slightly different to the used in \autoref{eq:stellar_mass}, we took
the corrected magnitudes in $B$ and $R$ from \cite{FreitasLemes2014}
due to their central wavelength are the closer to those of $g$ and $r$
filters. The induced error for this assumption is negligible given
that the difference between these wavelengths only contribute with the
continuous of spectra (see figure 5 in
\cite{FreitasLemes2014}). $\epsilon$ is a parameter that depends of
initial mass function with a value of $0.15$ for this case
(\cite{Bell2003}).

To estimate the halo mass of the MG we used the recipes presented in
\cite{Villa2015}, where an inverse procedure to \cite{Mo1998} model is
performed. In \cite{Mo1998} given a set of parameters, in a
$\Lambda$CMD cosmology, the virial velocity $V_{200}$, the
concentration $c$ and spin parameter $\lambda$ for a dark matter halo
hosting a stellar disk with fractions of mass and angular momentum
$m_d$ and $j_d$ relative to those of the halo, the scale length $R_d$
and the rotation curve $V_c$ are obtained. Conversely, in
\cite{Villa2015}, using distributions built with data from SDSS,
knowing $R_d$ and $V_c$ and assuming $m_d$ and $j_d$, many
realisations of the model are made, therefore, the most probable
combination of $V_{200}$, $c$ and $\lambda$ of a halo that could host
a particular disk with $R_d$, $V_c$, $m_d$ and $j_d$ is obtained. This
method works well for massive galaxies, however, it is not well
calibrated for low mass galaxies. The halo mass for SG was extracted
from rotation curves acquired by D.L. Ferreiro in august of 1999 with
the telescope of $2.15\:m$ in the Complejo Astron\'omico de la Sierra
de Leoncito (CASLEO).

We estimate the gas mass for MG from the ratios for $M_{HI}/M_*$
presented in \cite{Catinella2010}. To estimate the gas mass of SG we
used the estimated value with its H$_\alpha$ luminosity presented in
\cite{Ferreiro2008}. Finally, the bulge-to-disk mass ratio for both
galaxies was defined such that the stability of the disk is guaranteed
when they were simulated in isolation, this ensures that dynamical and
morphological changes in the galaxies are due only to its
interaction. In \autoref{tab:galaxy_properties} we summarise the
results of masses and scale lengths for all galactic components used
for the simulation of the system.

\subsection{The orbit}\label{subsec:orbits}

Observations gave us position in the $XY$ plane of each galaxy and
$z$-velocity for SG relative to MG. Then it is necessary to find its
relative $z$ position and the other two velocity components, $v_x$ and
$v_y$. In general, in order to reproduce the merger, we need to find
the orbit leading to the current status of the system. To obtain that
information, we used a two step approach. First a two body
approximation was used to explore the parameter space of orbital
parameters, and then an extended body was used to select candidate
orbits for the merger.

For the first step, assuming that both galaxies are point masses, we
systematically explored intervals in orbital parameters eccentricity
($\epsilon$), periastron ($q$), ($z$) position and the inclination of
the plane of the orbit ($i$).

In the two body approximation, a large set of orbits was produced
exploring the parameter space ($\epsilon, q, z$, $i$). Each orbit was
projected in the sky and it was accepted as a candidate orbit if it
was able to reproduce the observational constraints (observed position
and velocity of SG). Around 240000 candidate orbits came out of this
exploration.

In the second step we took the candidate orbits produced in the first
step and perform an integration in time of the orbit. For that, MG was
not considered as a point mass, instead, it was modelled as an extended
rigid body composed of twelve thousand static particles representing
the bulge, disk and dark matter halo. Namely, these particles are
static then the galactic components do not have dynamics, thus the
motion of MG is given by the dynamics of its centre of mass.  The mass
distributions used for each component are described in Section
\ref{sec:ics} and their masses as in Table
\ref{tab:galaxy_properties}. In this way we can account for the
effects of the gravitational torques acting on SG, that in this step
is still modelled as a point mass. The motion of the centre of mass of
MG and SG were integrated using a leapfrog integrator with adaptive
time step, while the force field acting on SG considered its
interaction with each particle in the realisation of MG. Starting from
the instant of observation a first integration backwards in time was
ran during a dynamical time (of MG), then an integration forward in
time was performed. Thus, if SG orbit passes again in position and
velocity close to the observational constraints this orbit is selected
as a potential orbit to run in a simulation. This reduced the number
of candidate orbits to around 12000 potential orbits.

As a last step, a minimisation of a chi-square estimator was made on
those 12000 orbits using as parameters the observational constraints
($x, y, v_z$). Thus we simulated AM 2229-735 taking the orbits with
the smallest values of chi-square. Finally, 54 low resolution (VLR)
simulations with different orbits were performed, from these
simulations we focus on the orbit that best reproduce the observations
of the system. Details of simulations are described in
\autoref{sec:simulations}.

\subsection{Initial conditions for simulations}
\label{sec:ics}

With all information obtained of AM 2229-735 for both, individual
galaxies as well as the orbit described by the galaxies in the system,
it is possible to build initial conditions of the numerical
realisations of the galaxies and then reproduce the interaction
between MG and SG. Initial conditions follow the prescription
presented in \cite{Springel2005}.  All relevant parameters for the
realisation of the particle distribution are summarised in Table
\ref{tab:galaxy_properties}. The spheroidal components, halo and bulge
follow a density profile of \cite{Hernquist1990}

\begin{equation}
  \rho(r)=\frac{M}{2\pi}\frac{a}{r(r+a)^3},
\end{equation}

\noindent where $M$ is the total mass of the component and $a$ is the
corresponding radial scale length. Gas and stellar disks, with the
same radial scale length, have exponential density profiles

\begin{equation}
\rho_{d}=\frac{M_{d}}{4\pi R_d^2z_o}\exp(-R/R_d) sech^2\left(\frac{z}{z_o}\right).
\end{equation} 

\noindent where $M_{d}$, $R_d$ and $z_o$ are the (stellar or gaseous)
disk mass, radial scale length and vertical scale length,
respectively.

 \begin{figure}
   \centering
   \includegraphics[scale=0.161,angle=270]{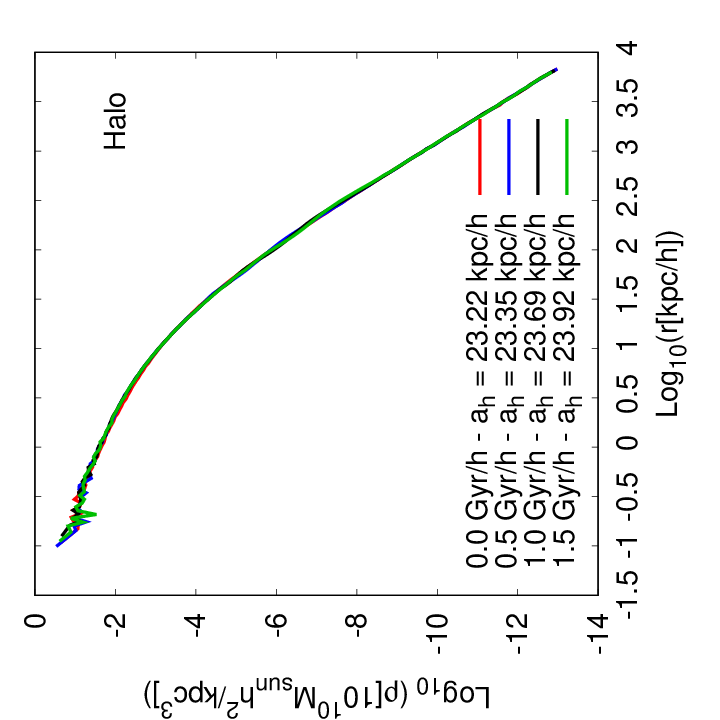}
   \includegraphics[scale=0.161,angle=270]{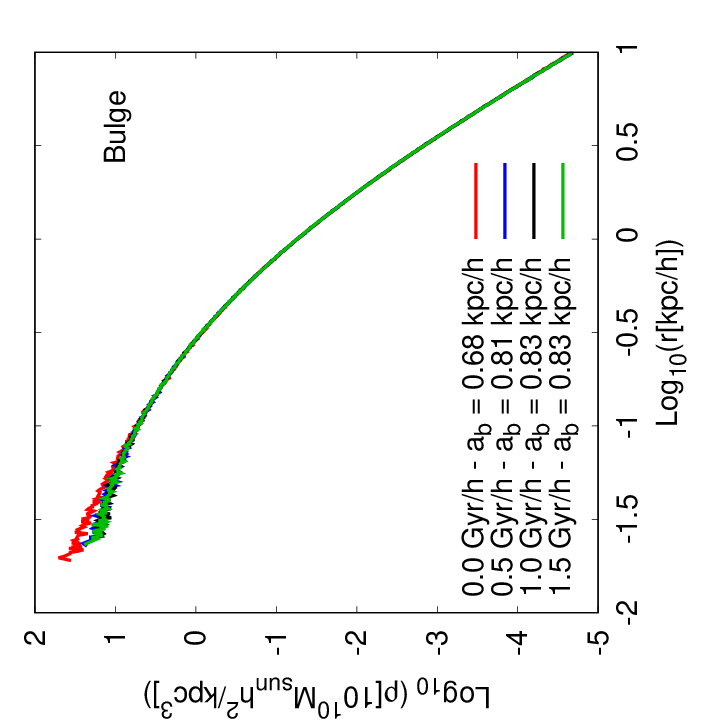}\\
   \includegraphics[scale=0.162,angle=270]{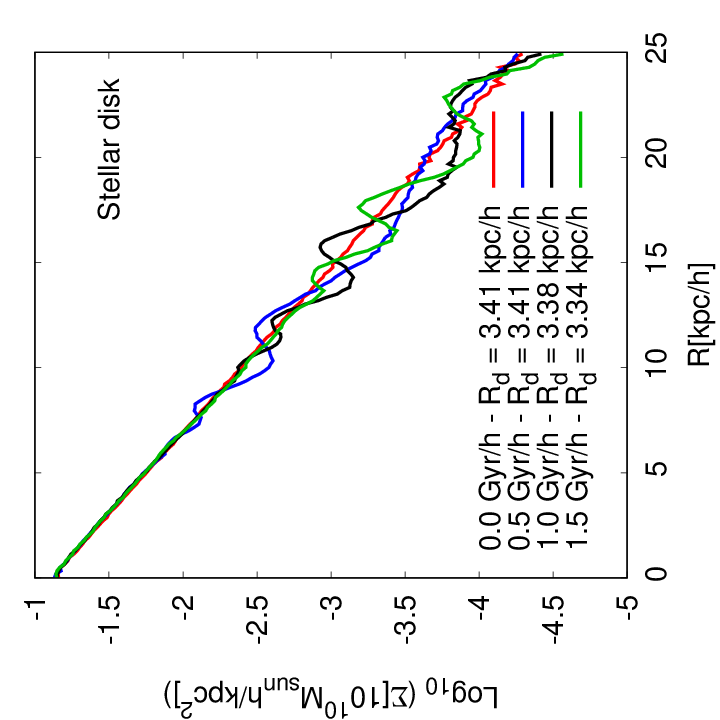}
   \includegraphics[scale=0.162,angle=270]{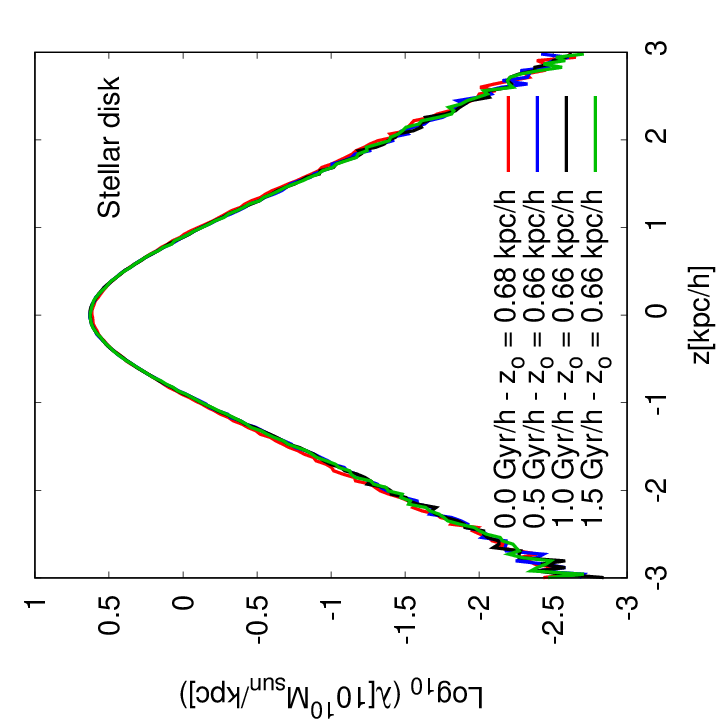}
   \caption{Density profiles for halo (top left panel), bulge (top
     right panel) and stellar disk (bottom panels) of main galaxy of AM
     2229-735. In all plots each line is the density profile in an
     specific time; red line in $0.0\:Gyr/h$, the blue line in
     $0.5\:Gyr/h$, the black line in $1.0\:Gyr/h$ and green line in
     $1.5\:Gyr/h$. For each time is indicated the scale length of
     profile. In bottom left panel and bottom right are the radial
     density profile and vertical density profile for stellar disk,
     respectively.}
      \label{fig:AM_relaxations}
  \end{figure}

Each galaxy was simulated in isolation for around $1.5\:Gyr/h$ to
test for stability of the initial conditions and to allow the model to
relax numerically. This way we ensure that any change observed in the
structure of the galaxies is induced by the merger not by any
instability induced by the initial conditions.

Figure \ref{fig:AM_relaxations} shows an example of stability check
for MG corresponding to high resolution simulation; each line represents the system
at a different time from 0 to $1.5\:Gyr/h$. Top left panel shows
the density profile for dark matter halo, this component does not
undergo significant changes in its structure, the shape of the density
profile is the same for all times and the value of the scale length is
almost constant. For the bulge (top right panel) its central density
decreases during the first $0.5\:Gyr/h$ while the scale length
grows to $0.8\:kpc/h$; After this relaxation the bulge's
structure remains stable in time. In the case of the stellar disk, the
radial structure (bottom left panel) shows full stability in the inner
region (within two times the radial scale lengths) all times. For the
outer regions density waves travel outwards from the disk, this
behaviour is typical in axis-symmetrical gravitational systems
initialised using the approach used in this work
\citep{Hernquist1993}, however after $1.5\:Gyr/h$ the waves
dissipate and equilibrium is guaranteed within more than four times
the initial scale length. The vertical structure of the stellar disk
remains almost unchanged for all times as is evident in bottom right
panel. All initial conditions used for MG and SG used in this work
have a similar behaviour when are simulated in isolation then the
interactions were configured using the galaxy models after $1.5\:Gyr/h$ of isolated numerical relaxation.

\section{Simulations}
\label{sec:simulations}

All simulations presented in this work were ran using the code Gadget2
\citep{Springel2005a}. It is a massively parallel TreeSPH code where
interactions of collisionless particles like stars and dark matter are
followed with a TreePM method while for the collisional fluid an
implementation of smoothed particle hydrodynamics (SPH) for an ideal
gas, that conserves energy and entropy in regions without dissipation
and fully adaptive smoothing lengths is used. The integration scheme
is a quasi-symplectic KDK leap-frog with adaptive individual
time-steps under a synchronisation scheme. Gadget2 uses a
parallelisation algorithm based on a space-filling curve getting high
flexibility with minimal implications in tree force errors.

\begin{table*}
\centering
\caption{Number of particles and smoothing lengths by galactic
  component for the three simulations of AM 2229-735. Columns 2 to 5
  correspond to main galaxy components and 6 to 9 to satellite galaxy
  components. Columns 10 and 11 are the smoothing lengths for dark
  matter and baryons components.}
\label{tab:galaxy_numbers}
\begin{tabular}{ l | cccc | cccc | cc }
\hline
Run & Halo MG & Disk MG & Bulge MG & Gas MG & Halo SG & Disk SG & Bulge SG & Gas SG & $\epsilon_{dm}$ (kpc/h) & $\epsilon_{b}$ (kpc/h)\\
\hline
LR & $3.05\times10^5$  & $1.23\times10^5$  & $5.32\times10^4$ & $1.59\times10^4$ &  $1.52\times10^5$ & $2.80\times10^4$  & $7.39\times10^3$  & $1.20\times10^4$  & 0.17 & 0.04  \\
MR  & $2.34\times10^6$ & $1.40\times10^6$ & $5.98\times10^5$ & $1.59\times10^5$ & $1.01\times10^5$ & $2.66\times10^4$ & $6.65\times10^3$  & $3.00\times10^4$ & 0.08 & 0.01  \\
HR  & $4.68\times10^6$ & $5.59\times10^6$ & $2.39\times10^6$ & $1.59\times10^6$ & $2.01\times10^5$ & $1.06\times10^5$ & $2.66\times10^4$ & $3.00\times10^5$ & 0.06 & 0.007 \\
\hline
\end{tabular}
\end{table*}

Our runs included basic hydrodynamics with radiative cooling for a
primordial mixture of hydrogen and helium \citep{Katz1996}, star
formation, galactic winds from supernovae feedback and chemical
enrichment, these models are described in \cite{Springel2003}. Here,
we annotate that although the radiative cooling model used here is not
metal dependent, making our simulations not completely
self-consistent, the metallicity acts only as a tracer of heavy
elements but without any impact on the dynamics of ISM.

To study numerical convergence, in this work have ran simulations at
three different resolutions, the low resolution (LR), medium
resolution (MR) and high resolution (HR); the number of
particles-per-component are summarised in
\autoref{tab:galaxy_numbers}. These numbers were tuned to make the
mass of each gas particle to take values of $1.0\times10^{3}$,
$1.0\times10^{4}$ and $1.0\times10^{5}$ $M_\odot/h$ in each
resolution. Masses for particles for the stellar disk and bulge were
always comparable to minimise two body relaxation effects. In the LR
simulations the number of particles for the satellite was modified to
increase the particle number of each galactic component in the stream
produced during the interaction. In Table \ref{tab:mass_resolutions}
we show the mass per particle for each simulation.

\begin{table}
\centering
\caption{Mass per particle in $M_{\odot}/h$ used in each resolution for
  gas, halo, disk and bulge galactic components.}
\label{tab:mass_resolutions}
\begin{tabular}{c|cccc}
\hline
Resolution & $m_{gas}$ & $m_{halo}$ & $m_{disk}$ & $m_{bulge}$\\
\hline
LR host & $1.0\times10^{5}$ & $2.3\times10^{6}$ & $4.5\times10^{5}$ & $4.5\times 10^{5}$\\
LR sat & $1.0\times10^{5}$ & $2.0\times10^{4}$ & $3.0\times10^{4}$ & $3.6\times 10^{4}$\\
MR & $1.0\times10^{4}$ & $3.0\times10^{5}$ & $4.0\times10^{4}$ & $4.0\times 10^{4}$\\
HR & $1.0\times10^{3}$ & $1.5\times10^{5}$ & $1.0\times10^{4}$ & $1.0\times 10^{4}$\\
\hline
\end{tabular}
\end{table}   

We performed 54 LR simulations whose orbits were extracted from the
final set of candidate orbits found with the procedure described in
\autoref{subsec:orbits}. The aim of these simulations was to find a
good orbit for the system AM 2229-735 such that the observational
constraints were well reproduced. The best orbit of this set was used
to run the MR and HR simulations. The results presented here about the
formation and evolution of the tidal stream produced during the
interaction corresponds to those analysed in HR simulation, while the
LR, MR are used to make a resolution study to determine the impact of
discretisation in the evolution of the system.

As it was described in \autoref{subsec:orbits} we know the trajectory
of the centres of mass of MG and SG between the observational point
and a point one dynamical time in the past. As each of this orbits is
elliptical and in all of them the observational point is achieved
after the first pass by the orbit periastron, then we selected the
apoastron just before that first periastron to start the
simulations. Finally, The running time of these LR simulations was
$10\:Gyr/h$ after that initial point and as each simulation
produced a trajectory different for the system, the orbit that
produced the morphology closer to that registered in the observations
of AM 2229-735 was chose to run the MR and HR simulations. The
behaviour of the selected orbit and how the morphology of the
interaction is reproduced will be described in
\autoref{subsec:reproduction}.

\section{Extracting the tidal streams from simulations}
\label{sec:stream}

With the aim to study the formation, evolution and properties of the
tidal stream produced during the interaction of AM 2229-735, it
becomes relevant to separate the particles forming this structure from
those belonging to the main and satellite galaxies. Thus, the physics
of the stream can be studied considering it as an independent
structure within the system AM 2229-735. As we will see in next
sections, our simulations show that this stream will form the polar
structure around the main galaxy.

We will assume that the stream will be formed mainly by particles
coming from the satellite galaxy. Thus, the tidal stream contains
mainly particles from each galactic component of the satellite: gas,
dark matter, disk and bulge stars and young stars formed in this
galaxy during the simulation. Thus we extracted the tidal stream
coming from each component separately to study its contribution in the
formation and evolution of the polar structure.

The tidal streams from collisionless particles were extracted using a
criterion of energy together with a criterion of local density.  For
each output snapshot of the simulation, the total mechanical energy of
particles of SG was computed relative to its centre of mass. Those
particles having a positive energy are considered unbound from SG
because they were stripped by MG. These particles can fall to the
centre or stay orbiting in the MG halo forming the tidal stream.

Then, to identify the particles belonging to the stream we used the
fact that the tidal stream is an overdensity in the region where it is
located. Using the software \textit{Enbid} \citep{enbid}, we estimate
the local density for particles of dark matter halo, in the initial
conditions of MG, and use it to build a curve of mean density as a
function of the position. Then, for each snapshot of the simulation we
calculated the local density for all particles that were labelled as
unbound from SG, thus we define that a particle belongs to the tidal
stream if its local density is larger than the local mean density of
the dark matter halo of the MG at the position of the particle.

With the previous procedure we can identify the collisionless
components in the stream. As an example we show in Figures
\ref{fig:XY_HR_better_feedback_posiciones} and
\ref{fig:XZ_HR_better_feedback_posiciones} the X'Y' and X'Z'
projections in a plane oriented with the angular moment of the stream,
the panels labeled with dark-matter, disk stars, bulge stars and young
stars depict the stream subtracted from the simulation at $0.5$,
$1.0$, $2.0$ and$3.0\:Gyr/h$ respectively. The plots clearly show that
this method is successful to find the stream of the collisionless
components.  However, in order to avoid the inclusion of particles
from the MG (new stars or gas particles) we reject from the stream any
particle that is at a distance, measured from MG centre, smaller than
$3.5$ times the scale lengths of the MG. In this way we keep all our
attention on the material deposited in the stream from the satellite
galaxy only.

Gas may be diluted in the galactic halo because heating induced by
different physical mechanisms. This hot atmosphere may complicate the
identification of cold gas gravitationaly bound to the stream. In
order to find the gas belonging to the stream, we will focus our
attention on the gas that is distributed alongside with the
collisionless particles. Then we define the gas in the tidal stream as
these gas particles inside the volume defined by the stream of
collisionless particles forming the tidal stream.

In order to do so, once the stream defined by collisionless particles
is identified, we use the particle distribution (in such a stream) to
build a Delaunay tessellation.  The tessellation will define the
volume enclosing the stream. Any gas particle inside any of the
tetrahedron defined by the collisionless particle distribution should
also belong to the stream.

In the panels at the left in Figures
\ref{fig:XY_HR_better_feedback_posiciones} and
\ref{fig:XZ_HR_better_feedback_posiciones} we show the X'Y' and X'Z'
projections of gas stream at $0.5$, $1.0$, $2.0$ and$3.0\:Gyr/h$ from
top to bottom respectively. As it can be seen in the figures, the
method is successful to identify the gas inside the stream. Note,
however, that this gas stream will have particles of gas coming from
tidal stripping or gas that was there in the hot gas halo of MG.

\section{Results and discussion}
\label{sec:results}

Now we present the results of the simulations that we use to study the
evolution of AM 2229-735 to get clues about its fate and to study if
the formation of a polar structure is possible in this system, and if
it is the case, what are its structural properties.

\subsection{General aspects of the evolution and reproduction of observables}
\label{subsec:reproduction}

First we show the system global features depicted in our simulations
generated by the galaxies following the orbit that best reproduce the
observed constraints, these characteristics were studied with the HR
simulation that ran during $3\:Gyr/h$.

 \begin{figure}
   \centering
   \includegraphics[scale=0.16,angle=270]{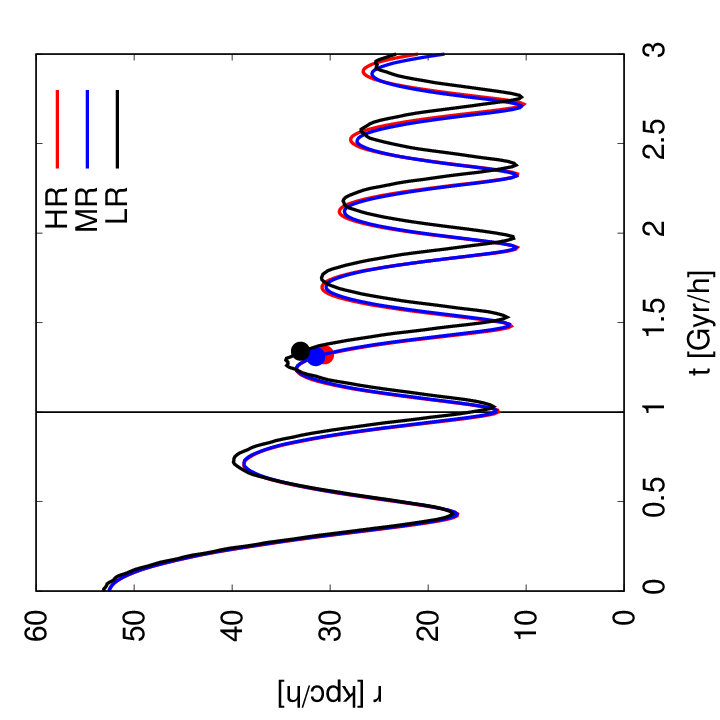}
   \includegraphics[scale=0.16,angle=270]{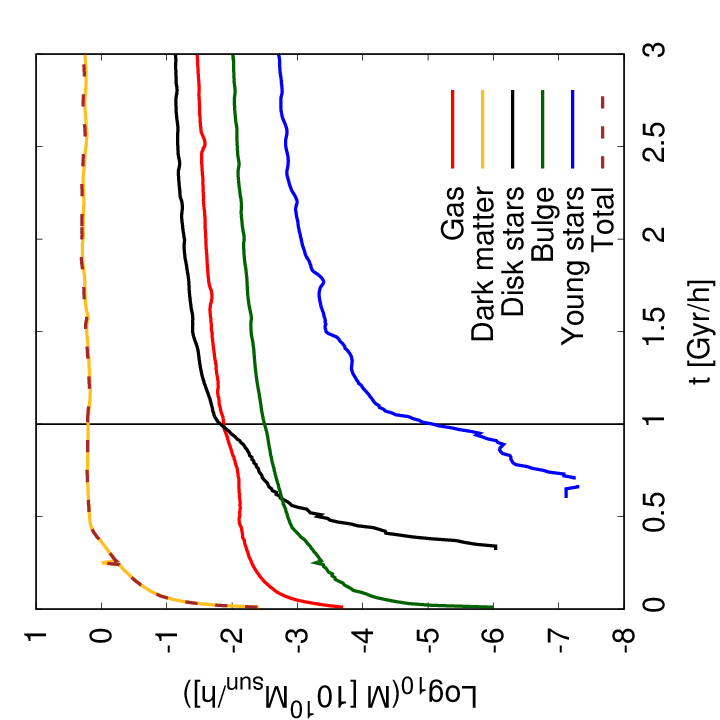}
   \caption{Left panel is distance to main galaxy centre as a function
     of time. The red line for HR, blue line for MR and black line for
     LR resolutions. The points on each line are the \textit{match
       point} for each simulation. The right panel is the mass found in
     tidal stream from each galactic component as a function of
     time. The colours for the mass of each component are: red for gas,
     yellow for dark matter, black for disk stars, green for bulge,
     blue for young stars. The red dashed line is the total mass of the
     stream.}
   \label{fig:AM_orbit}
 \end{figure}

The first thing we have to do is to be able to track the satellite
orbit during the simulation. We do so by tracking the most bound
particles in the satellite. We first identify the set of the most
bound five hundred particles at $t=0$ (relative to the centre of mass
of SG), and then track that set of particles across all snapshots
tracing the orbit of the satellite's core while the merger happens.

\autoref{fig:AM_orbit} (left panel) shows the satellite distance to
the MG's centre as a function of time, the red line corresponds to the
SG distance in simulation HR. The satellite begins a quasi periodic
orbit with regular crosses by periastron each $\sim 0.4, 0.5$
$Gyr/h$. This closed orbit changes in time due to the effects of
dynamical friction and tidal stripping. An important aspect to note is
that after $3\:Gyr/h$ a complete merger does not happen despite
satellite mass is decreasing at all times during the merger. SG's core
describes an orbit revolving around the main galaxy, the energy of
this orbit is such that the satellite is not swallowed by the main
galaxy and does not transfer all its material to it. The red
\textit{match point} shown in the left panel of \autoref{fig:AM_orbit}
marks the time in the simulation where this reproduces the
observational constraints and the current state of the merger
morphology. This matching with observations happens after the
satellite passes by second apoastron at time $t=1.32$ $Gyr/h$
after starting the simulation.

 \begin{figure}
   \centering
   \includegraphics[scale=0.35]{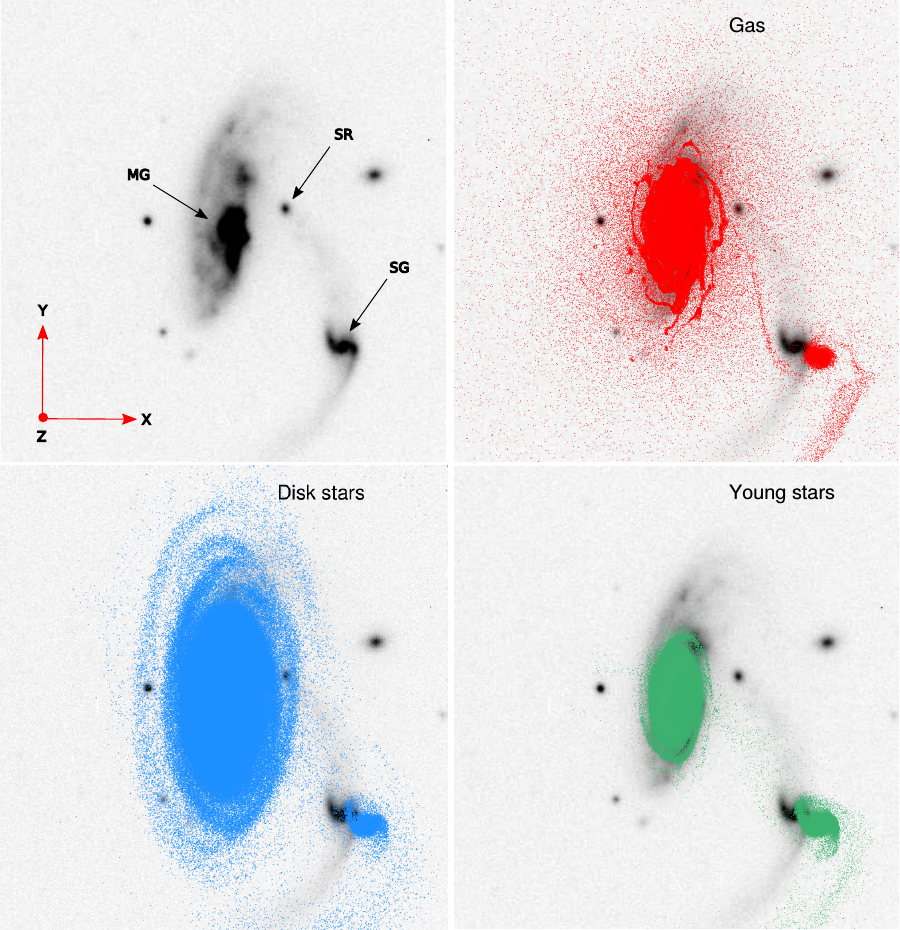}
   \caption{XY projections for some galactic components of both
     galaxies at match point at $\sim1.32\:Gyr$ after starting the
     simulation and the passage through the second periastron, the
     coordinates system and scale are the same of \autoref{fig:AM}.  In
     left-top panel is the image of AM 2229-735 where MG, SG and SR
     point to the centres of main galaxy, satellite galaxy and special
     region respectively as is explained in
     \autoref{sec:observations}. This image is used as background for
     the others panels where are superimposed points corresponding to
     gas (right-top), stellar disk (left-bottom) and young stars
     (right-bottom).}
   \label{fig:AM_superposicion}
 \end{figure}

In \autoref{fig:AM_superposicion} we show the XY projections of
different galactic components in the simulation for both galaxies at
the match point overploted on the observed system. In left-top panel
is the image of AM 2229-735, this image is used as background for the
other panels where are superimposed points corresponding to gas
(right-top), stellar disk (left-bottom) and young stars
(right-bottom). The stellar components display arms with a shape very
close to observations, the main galaxy disk developed a large arm
after the passage of the satellite that at match point has a position
very close to observed one. The same happens with the spiral arms
exhibited by the satellite, they are developed during the interaction
and have a shape and orientation similar to those found in the
observations. Additionally, all galactic components show a bridge of
material between galaxies at this time instant, this result is in
agreement with observations where gas, old and young stars content
were detected in the tidal streams of the system \citep{Ferreiro2008,
  FreitasLemes2014}. Finally, no structure like the especial region
(SR) is formed in our simulations, this leads us to think that maybe
that region is a field object and does not belong to the system AM
2229-735.

As it is observed in Figure \ref{fig:AM_superposicion}, the position
of the satellite and tidal stream do not match perfectly, this is
expected by us because the procedures described in
\autoref{subsec:orbits} only gives a set of possible orbits for the
system, then the orbit selected is a good orbit to reproduce the main
morphological features of the interaction but clearly is not the exact
orbit for AM 2229-735. Nonetheless, that we achieve to reproduce the
main morphological characteristics is a sign that the procedures
presented in \autoref{subsec:orbits} are useful to reconstruct the
orbit of the system. As the procedures used to get the possible orbits
for AM 2229-735 are independent of the system, these can be used for
any other observed minor merger provided that the observational
constraints of $x, y, v_z$, galaxy masses and orientation angles for
the galactic disks are known, and assuming that the merger is not in a
very advanced state of evolution.

 \begin{figure}
   \centering
   \includegraphics[scale=0.2]{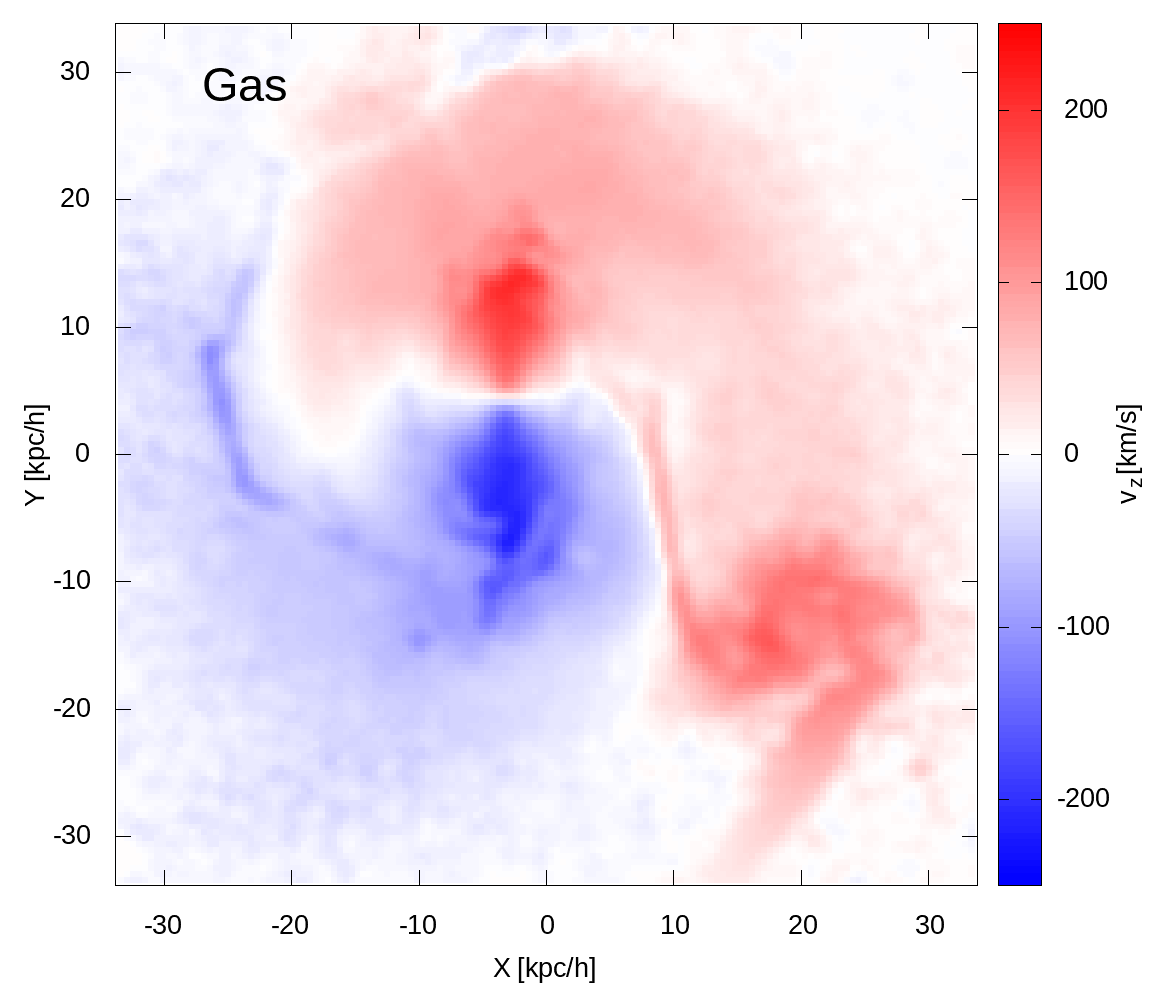}
   \includegraphics[scale=0.2]{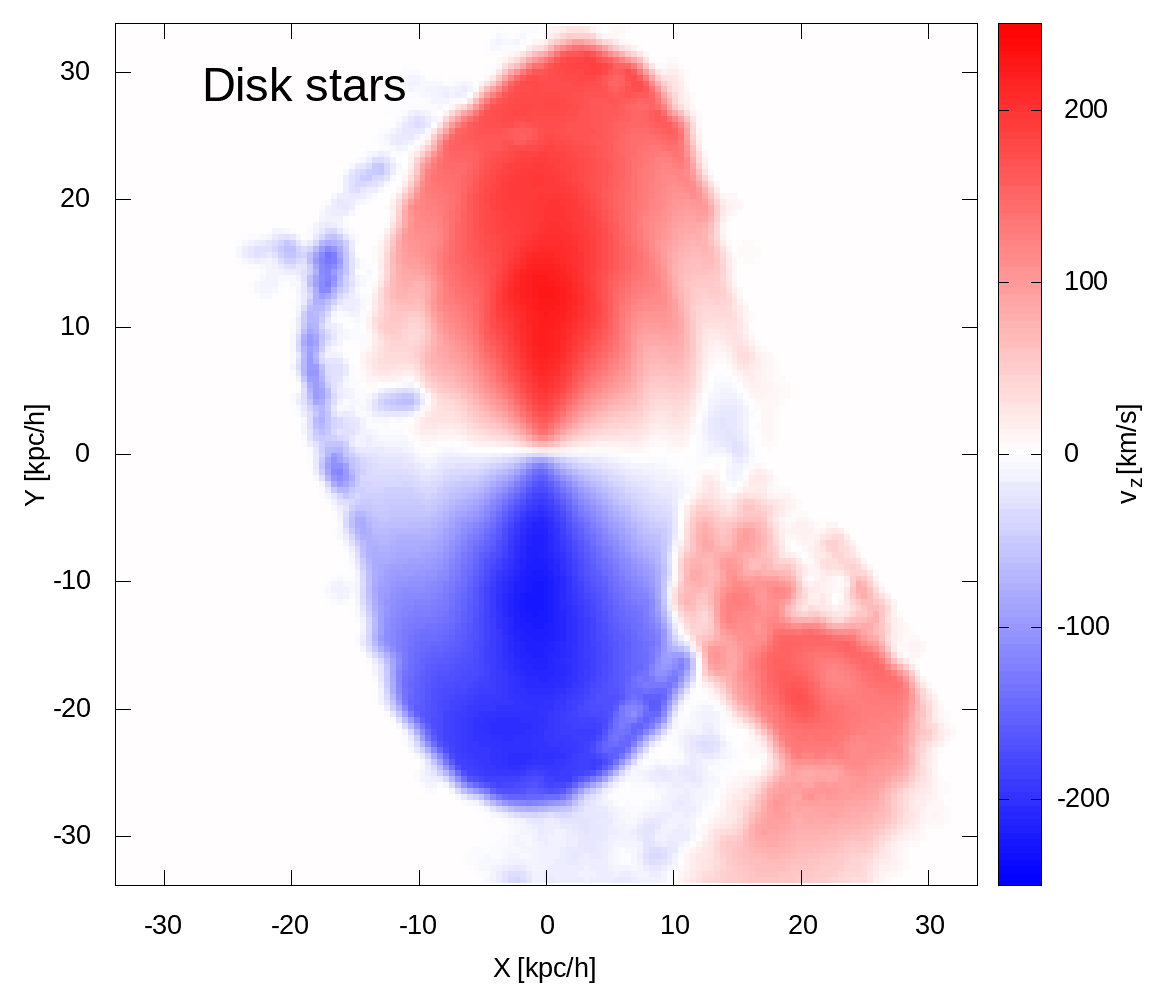}
   \caption{Velocity field along the line of sight for stream particles
     from gas (top-panel) and disk stars (bottom-panel). These plots
     show the mean velocity computed in cells of
     $0.46\times0.46\:(kpc/h)^2$. Each field is over-ploted on top of
     the XY projection of disk stars or gas in the simulation at mach
     point.}
   \label{fig:AM_velocities}
 \end{figure}

Besides the appearance of the merger, depicted in figure
\autoref{fig:AM_superposicion}, one could be interested in the
observed velocity field of the system. In order to approach what one
could observe if were able to observe the full velocity field of the
system, in Figure \ref{fig:AM_velocities} we draw the velocity field
projected along the line of sight for the particles in the stream from
gas (top-panel) and disk stars (bottom-panel). These plots show the
mean velocity computed in cells of $0.46\times0.46\:(kpc/h)^2$. Each
field is over-ploted on top of the the XY projection of disk stars or
gas in the simulation at mach point. Here we can see the range of
velocities of the material forming the stream that finally will form
the polar structure. These values could be compared with observations
of tidal tails observed in minor mergers.

\subsection{The formation of the polar structure of AM 2229-735}
\label{subsec:polar_structure_results}

As we mentioned before, the satellite looses mass while orbiting the
main galaxy, this mass is both dark matter and baryonic particles that
originally belong to the SG but that are tidally stripped falling to
MG forming a tidal stream around it. The right panel in
\autoref{fig:AM_orbit} shows the mass of the different components in
the tidal stream formed by particles stripped from the satellite at
each time. Each curve corresponds to the total mass stripped from each
component of SG that forms the stream. Note that the tidal stream is
formed with particles from all galactic components: dark matter halo
(yellow line), gas (red line), bulge (green line), stellar disk (black
line) and young stars (blue line), being the dark matter its more
massive component. We found that at $3\:Gyr/h$ the tidal stream
contains the $\sim 59 \%$ of the initial dark matter halo mass from
SG.  We found also that the stream contains $\sim 4.3 \%$ of the old
stars, $\sim 0.1 \%$ of young stars and $\sim 1.8 \%$ of gas from the
initial satellite mass.

As it can be seen from \autoref{fig:AM_orbit}, considering the
contribution from all components, the stream gets almost all its mass
after the second passage by the periastron at $\sim 1$
$Gyr/h$. After that, the increase in mass is not significant. Of
particular interest, is the contribution of mass of young stars in the
stream after $\sim 0.6\:Gyr/h$. These particles were created by
star formation processes in the satellite disk and then stripped to
the stream. No formation of new stars in the stream is observed in our
simulations.

 \begin{figure*}
   \centering
   \includegraphics[scale=0.195,angle=0]{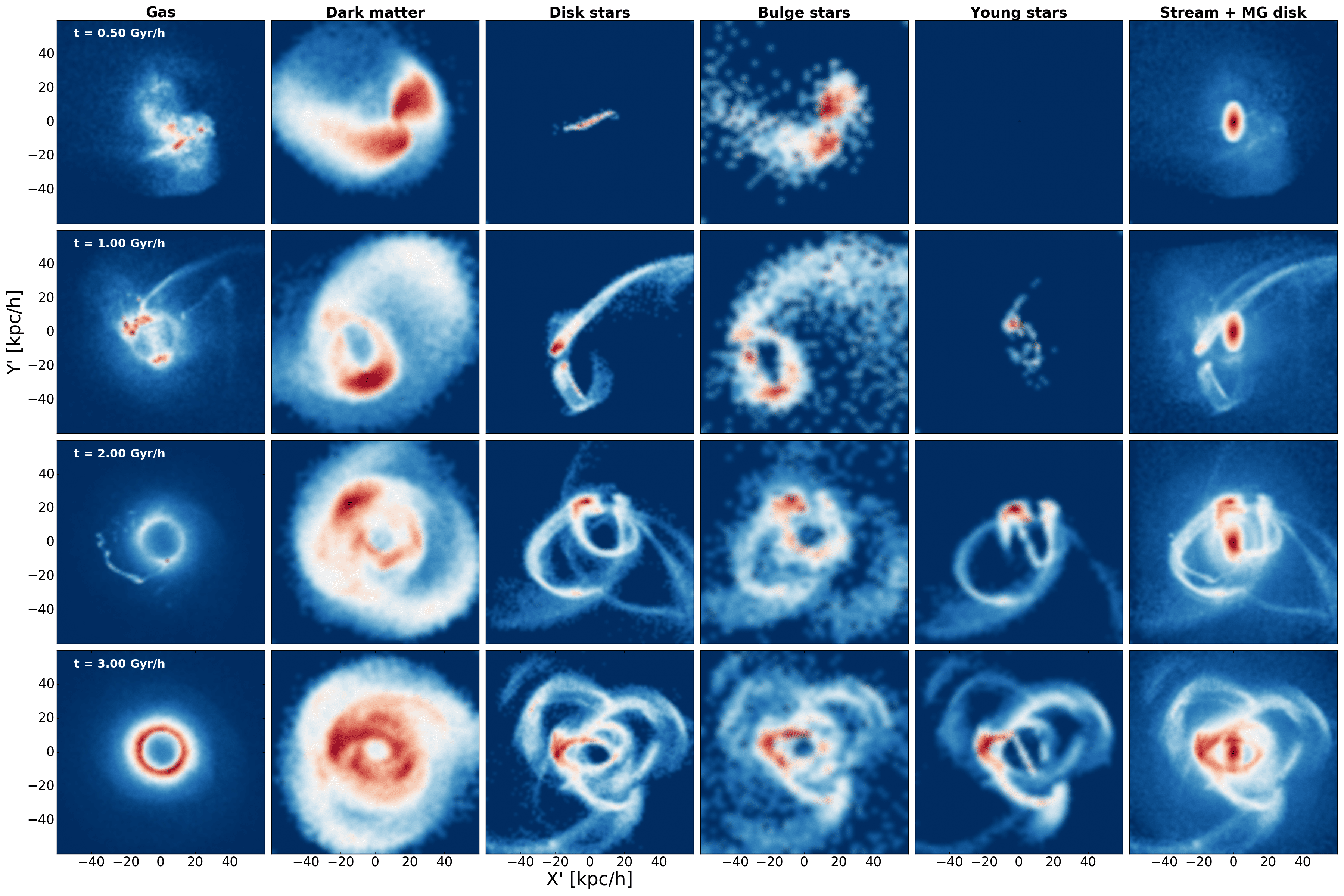}
   \caption{X'Y' projections for particles detected in stream coming
     from different galactic components of satellite, from left to
     right, each column are the projections for stream particles from
     gas, dark matter, stellar disk, bulge, young stars and, in the
     last column appear stars and gas stream joint to MG stellar disk
     particles. From top to bottom are the projection for times $0.5$,
     $1.0$, $2.0$ and $3.0$ $Gyr/h$, respectively.}
   \label{fig:XY_HR_better_feedback_posiciones}
 \end{figure*}

 \begin{figure*}
   \centering
   \includegraphics[scale=0.195,angle=0]{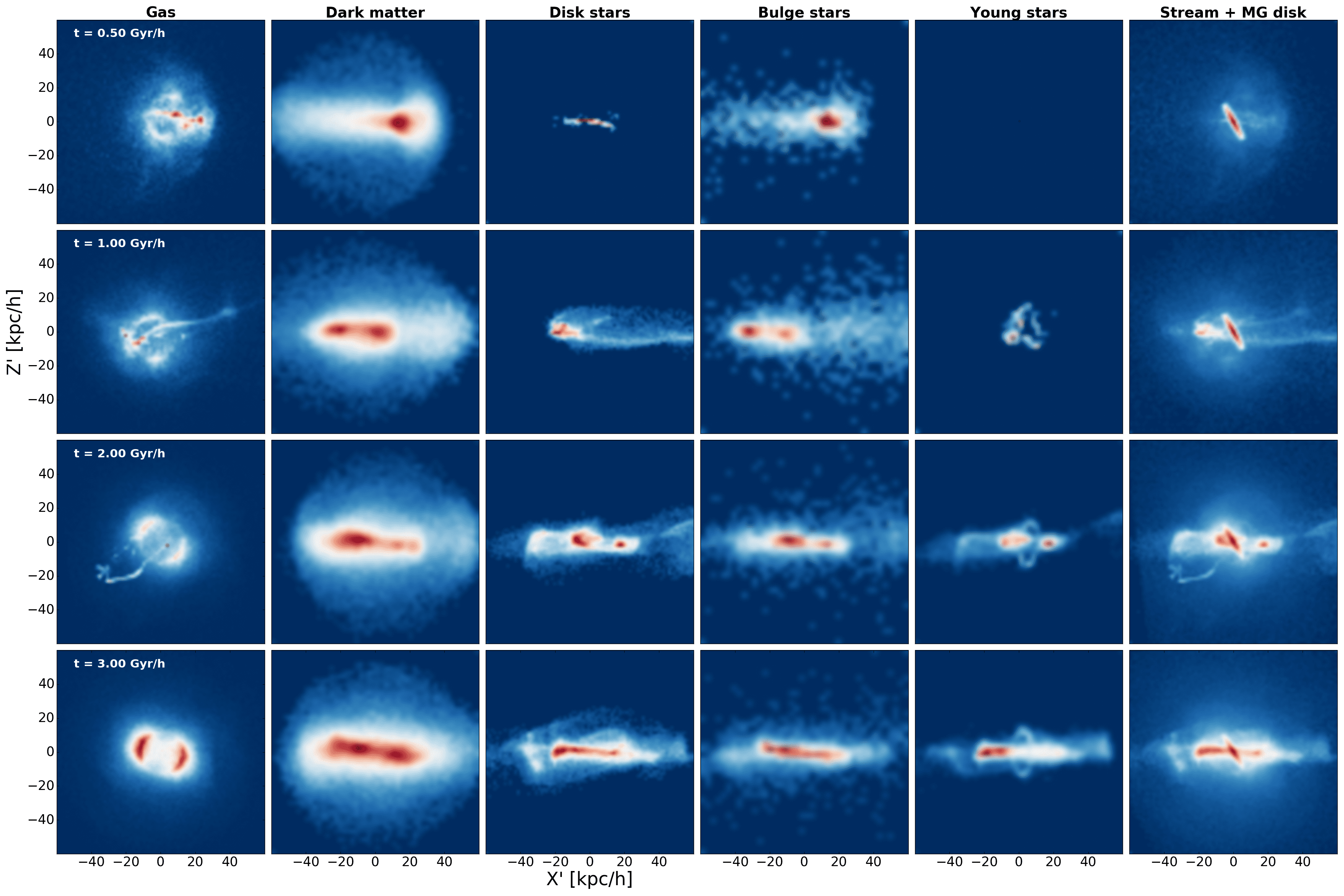}
   \caption{X'Z' projections for particles detected in stream coming
     from different galactic components of satellite, from left to
     right, each column are the projections for stream particles from
     gas, dark matter, stellar disk, bulge, young stars and, in the
     last column appear stars and gas stream joint to MG stellar disk
     particles. From top to bottom are the projection for times $0.5$,
     $1.0$, $2.0$ and $3.0$ $Gyr/h$, respectively.}
   \label{fig:XZ_HR_better_feedback_posiciones}
 \end{figure*}

Now, let's see how the mass distribution of the tidal stream forms a
polar structure around the disk of the main galaxy as it was suggested
in \cite{FreitasLemes2014}. For that, in Figures
\ref{fig:XY_HR_better_feedback_posiciones} and
\ref{fig:XZ_HR_better_feedback_posiciones}, we plotted projections
$X'Y'$ and $X'Z'$ respectively, of particles forming the stream
separated by the galactic component where they were originally bound
in the satellite galaxy. The figures are presented in the coordinates
$(x',y',z')$, with centre at the centre of mass of MG, of the
particles in the stream after rotations such that the angular momentum
of the stream points along the $z'$ axis, thus, we are showing its
face-on and edge-on views. In the panels of these figures we see, from
top to bottom, the formation and evolution of the stream during the
interaction as visualised for times $0.5$, $1.0$, $2.0$ and $3.0$
$Gyr/h$ after the start of the simulation. For these images the colour
code is scaled by density and we show the tidal stream by components,
from left to right, gas, dark matter, stellar disk, bulge, young stars
and the stars and gas stream together to MG stellar disk particles
(for a reference).

Gas mass is present in the stream by two processes; first, a fraction
of gas mass in the galactic disks of both galaxies is expelled from
them due to supernovae feedback and winds. In our simulations we used
a strong galactic wind isotropically distributed around disks and its
mass depends on the efficiency of production \citep{Springel2003}. The
second process is gas stripped out from SG forming gas tails that are
observed in the panels of first column of Figures
\ref{fig:XY_HR_better_feedback_posiciones} and
\ref{fig:XZ_HR_better_feedback_posiciones}. One of those tails formed
after second passage by periastron is part of the bridge shown in
\autoref{fig:AM_superposicion} and it is currently observed. Part of
those particles settled down in to a rosette that together with the
other components are forming the polar structure.

 \begin{figure*}
   \centering
   \includegraphics[scale=0.179,angle=0]{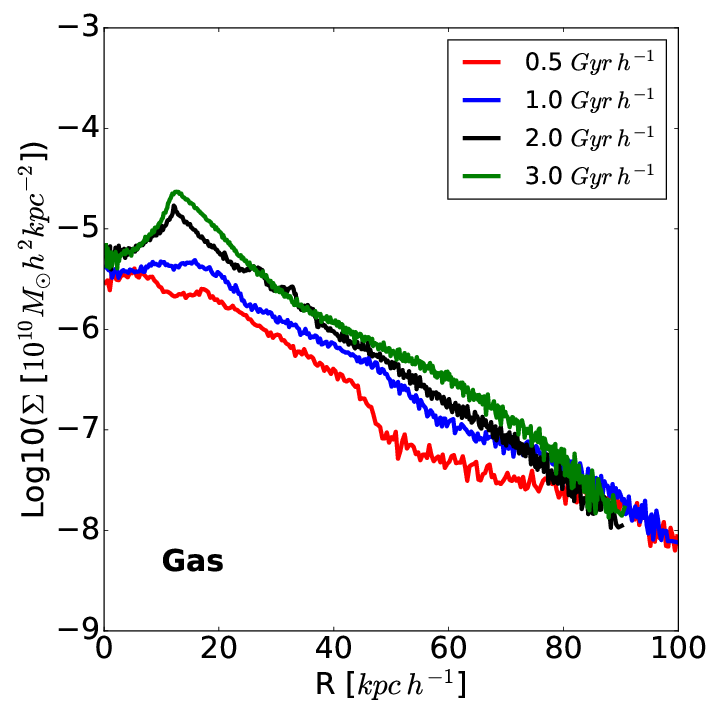}
   \includegraphics[scale=0.179,angle=0]{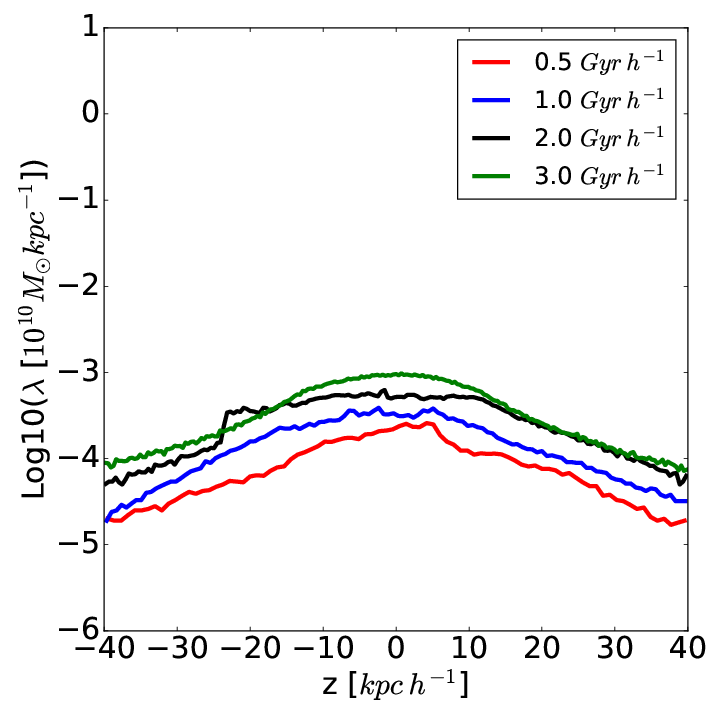}
   \includegraphics[scale=0.179,angle=-0]{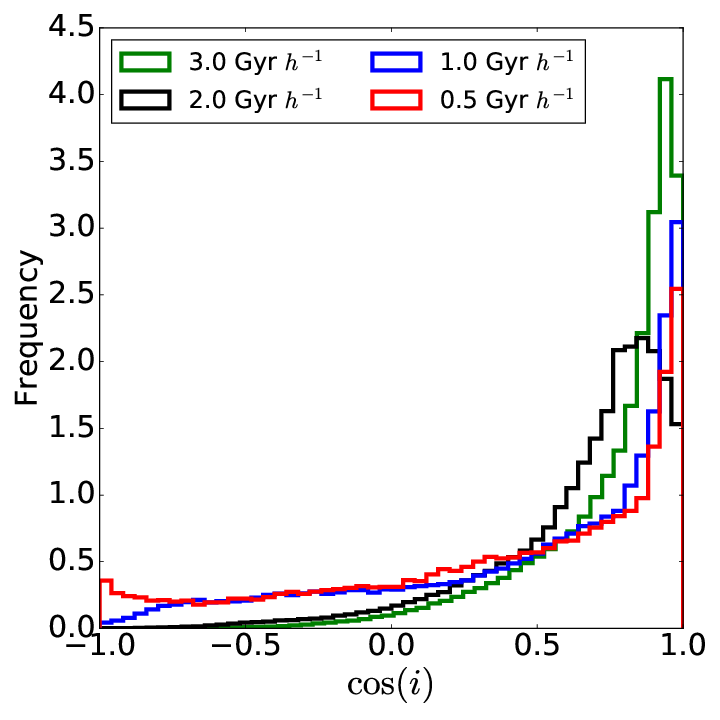}
   \includegraphics[scale=0.179,angle=0]{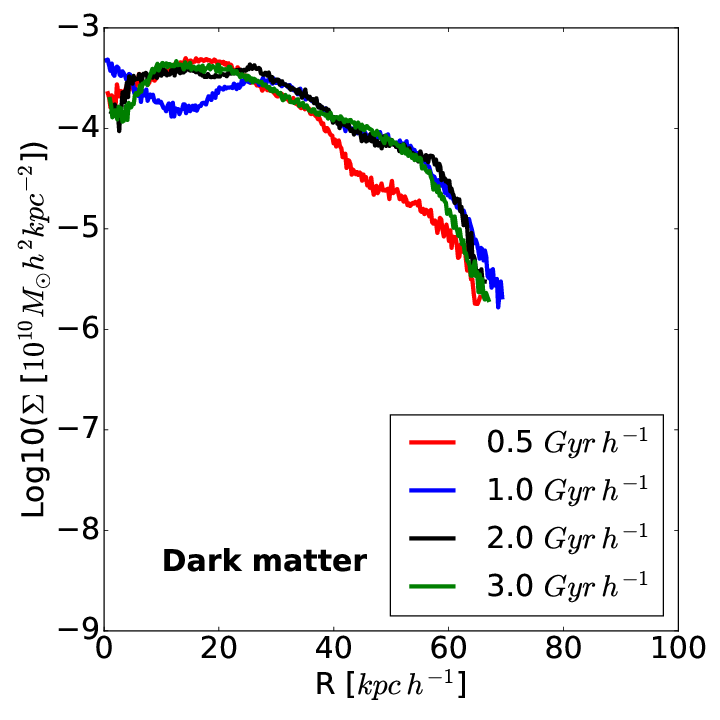}
   \includegraphics[scale=0.179,angle=0]{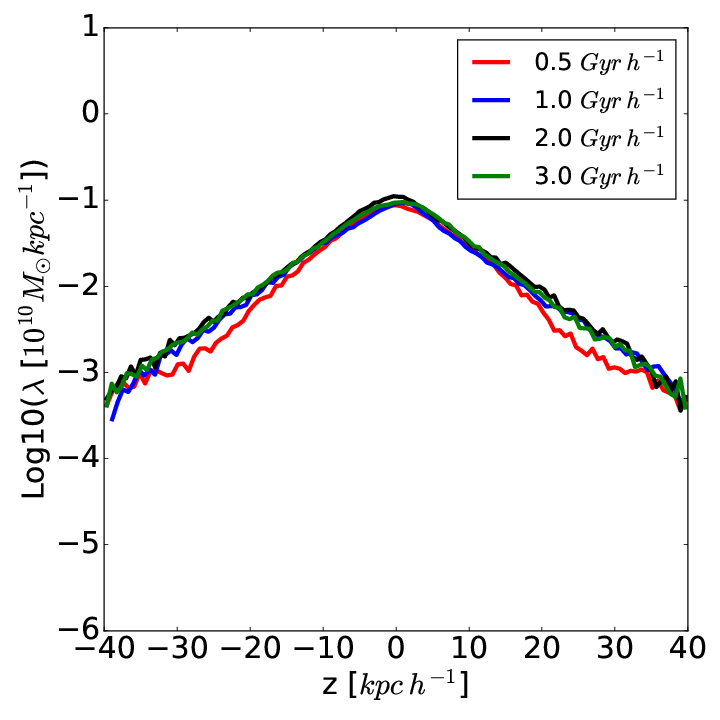}
   \includegraphics[scale=0.179,angle=0]{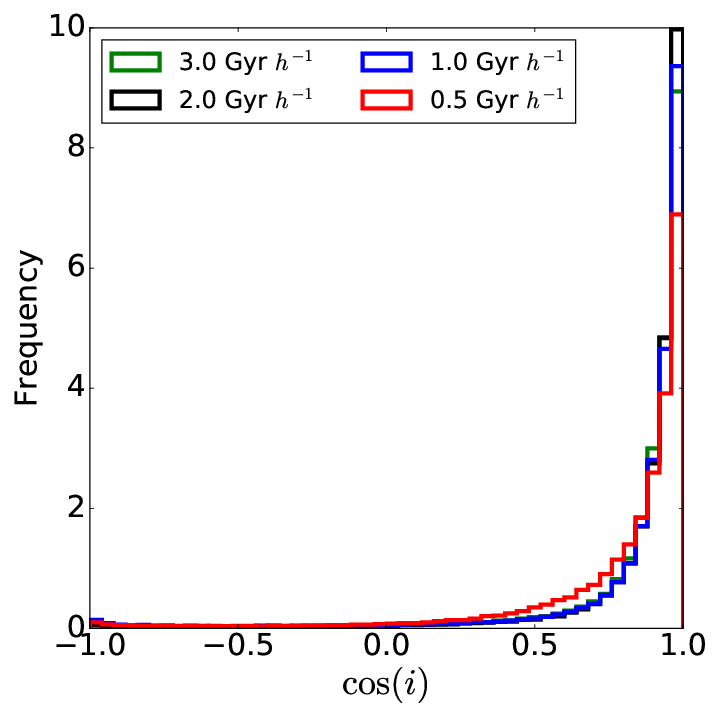}
   \includegraphics[scale=0.179,angle=0]{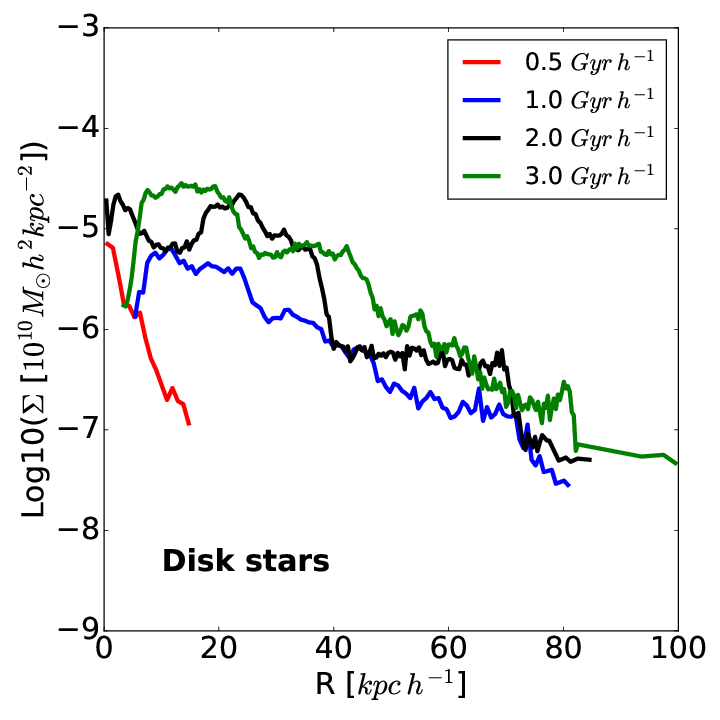}
   \includegraphics[scale=0.179,angle=0]{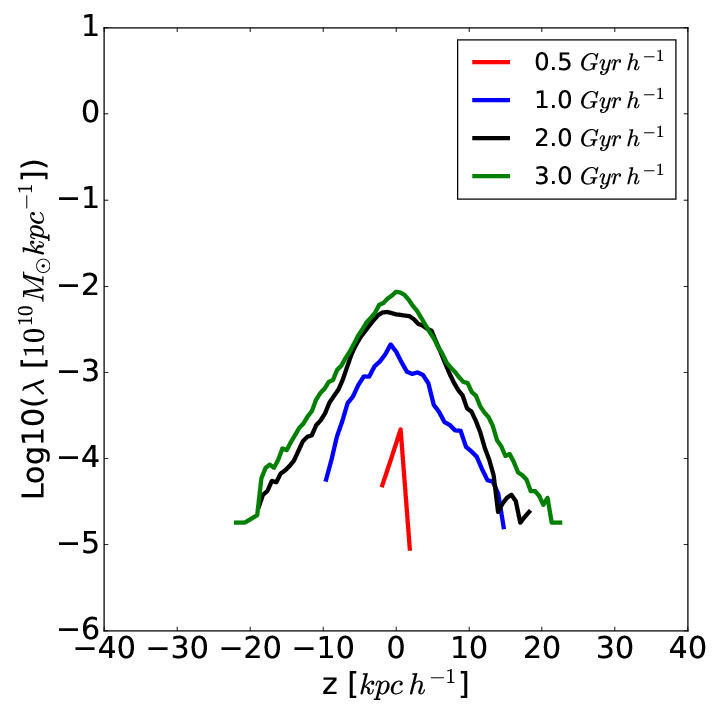}
   \includegraphics[scale=0.179,angle=0]{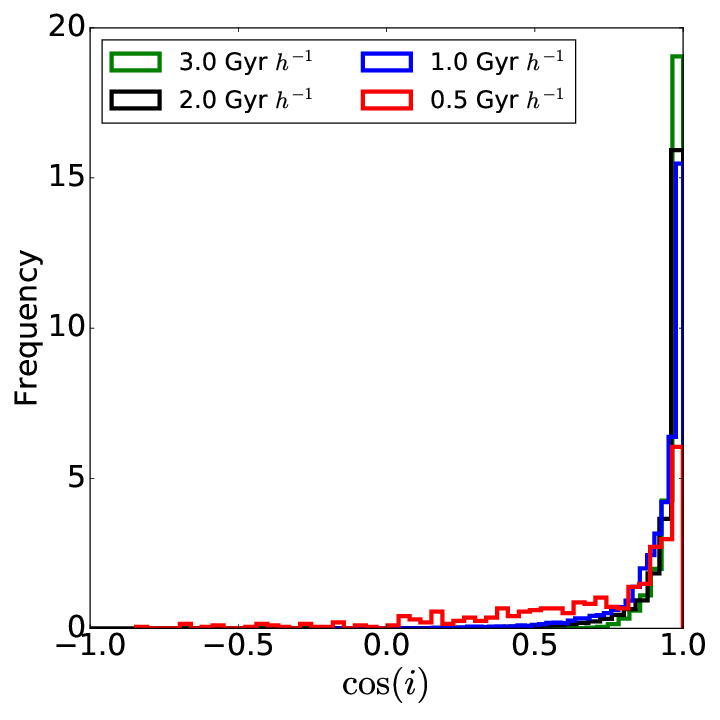}
   \includegraphics[scale=0.179,angle=0]{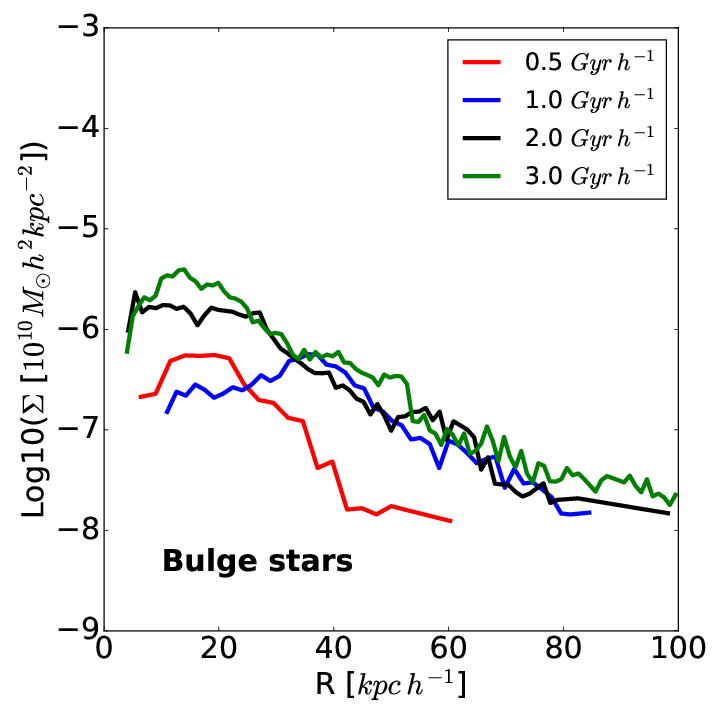}
   \includegraphics[scale=0.179,angle=0]{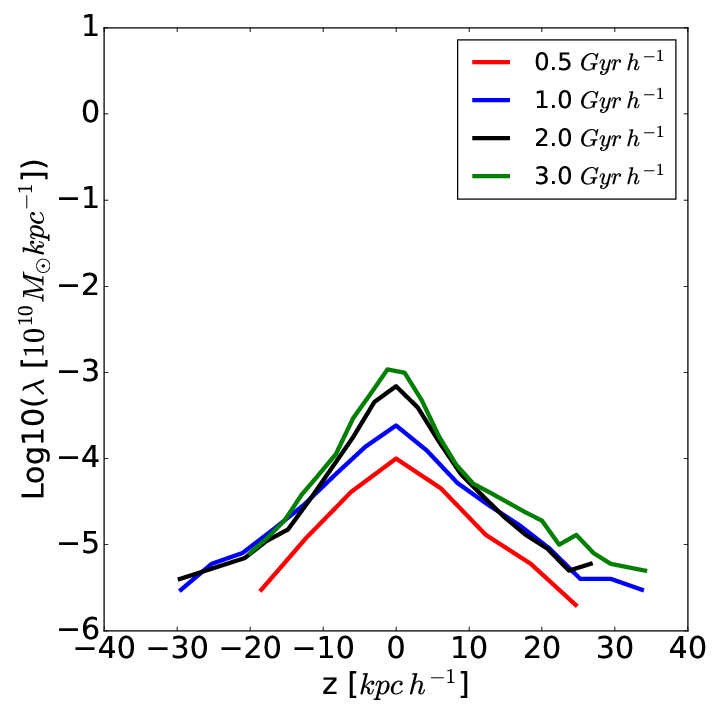}
   \includegraphics[scale=0.179,angle=0]{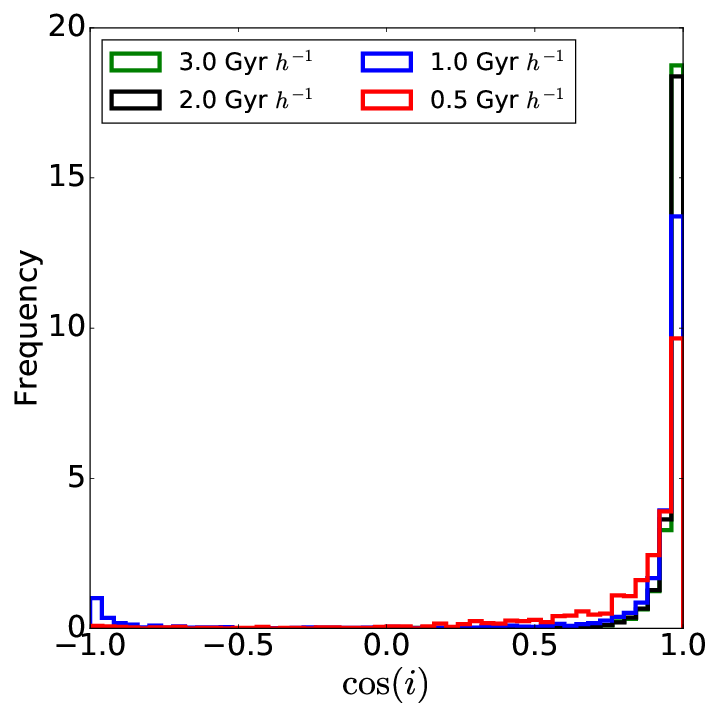}
   \includegraphics[scale=0.179,angle=0]{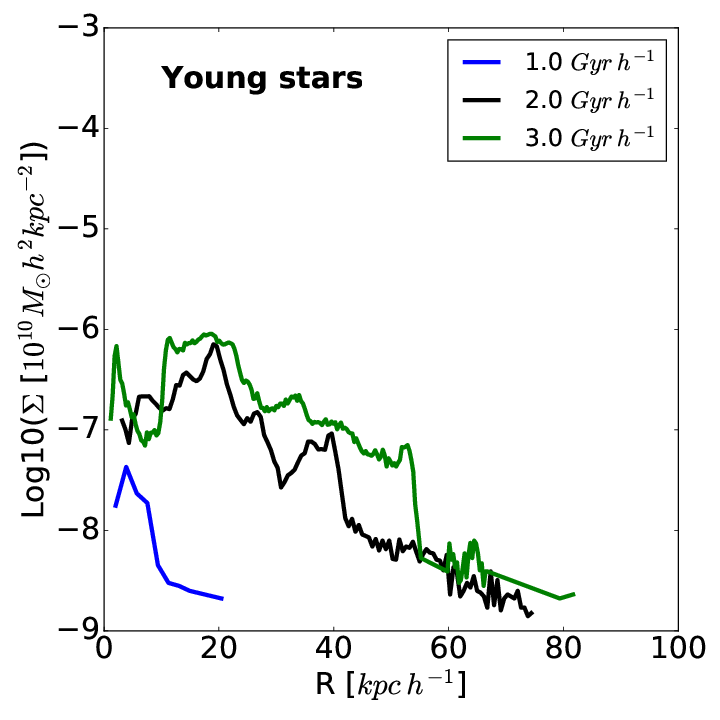}
   \includegraphics[scale=0.179,angle=0]{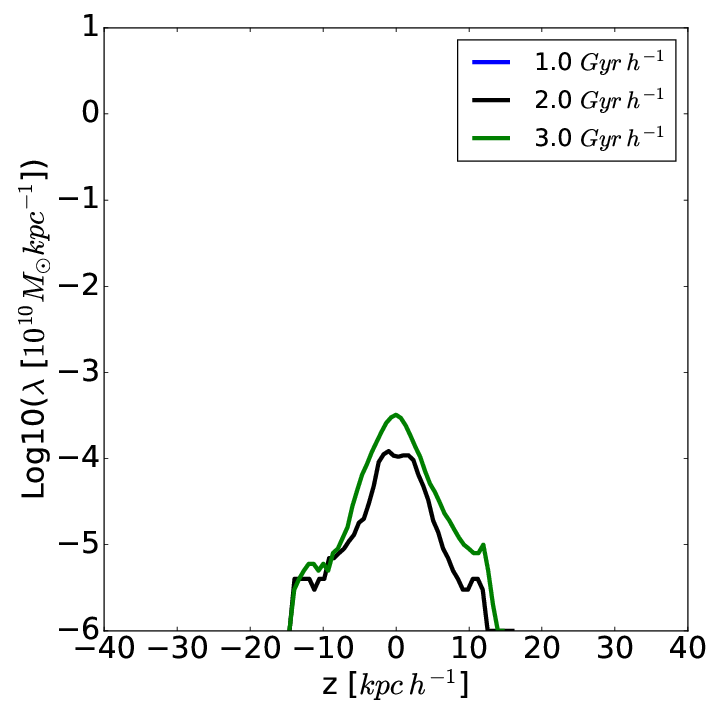}
   \includegraphics[scale=0.179,angle=0]{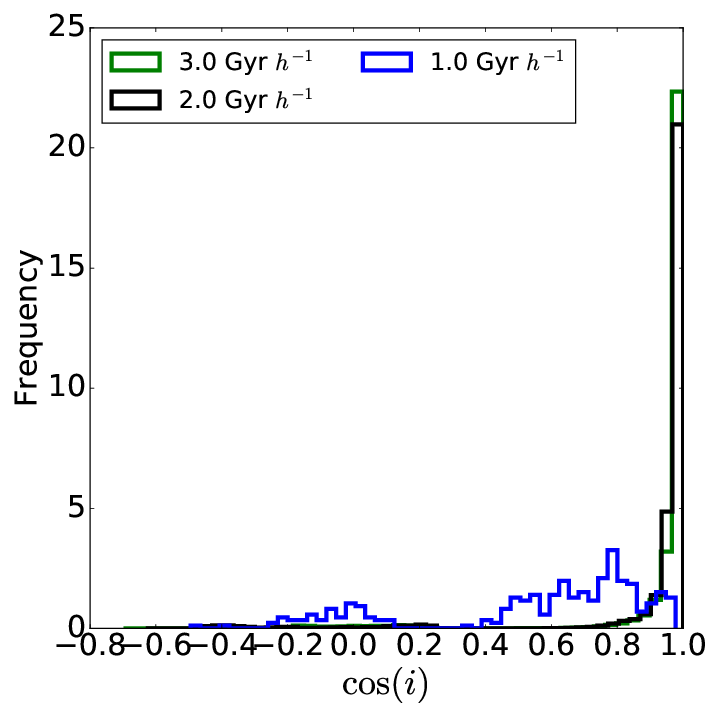}
 \caption{From left to right, are the surface density profile, vertical
   density profile and distribution of cosine of inclinations for
   particles from each galactic component at times $0.5$, $1.0$, $2.0$
   and $3.0\:Gyr/h$, from top to bottom, gas, dark matter,
   stellar disk, bulge and young stars.}
 \label{fig:evolution_stream_HR_better_feedback}
 \end{figure*}

The particles of dark matter, stellar disk and bulge form a rosette
around the main galaxy, this structure begins its formation with the
passage of the satellite by first periastron and at a time of around
$2$ $Gyr/h$ its final shape is already defined. A very notorious
characteristic of this structure is its planar distribution, following
these components in time it is clear that while the rosette formation
happens the material accommodates in a plane that in our projections
matches with the $X'Y'$ plane. This is due to the fact that the
orbital initial conditions are such that the stream inherits part of
the orbital angular momentum of the satellite conserving its direction
producing a planar structure.

The young stars, that as we described before formed in the satellite
gaseous disk, were stripped and appear in the tidal stream after $0.6$
$Gyr/h$ with a distribution similar to the distribution assumed
by the stars from the stellar disk. Thereby, a polar structure with
mass of dark matter, gas and old and young stars forming a rosette in
a plane is produced during the interaction with an inclination almost
perpendicular to the main galaxy disk. In panels of the last column of
Figures \ref{fig:XY_HR_better_feedback_posiciones} -
\ref{fig:XZ_HR_better_feedback_posiciones} we plot the projections of
stellar disk of MG together to the polar structure formed with the
stars and gas from satellite galaxy. The projection of the polar
structure and the stellar disk of MG at $3.0\:Gyr/h$ depicts an
appearance similar to those galaxies classified as PRGs, for example,
NGC 4650.

To analyse a bit more this polar structure we study its mass
distribution using surface and vertical density profiles in the
coordinate system $(x',y',z')$. Additionally, we determine the
inclinations of the orbits of these particles relative to the galactic
disk of MG. For that, we compute the angular momentum of each particle
relative to the stellar disk of the main galaxy. Then, we compute the
inclinations as the angle between the angular momentum of each
particle and the angular momentum of the stellar disk of
MG. \autoref{fig:evolution_stream_HR_better_feedback} shows, from left
to right, the surface density profile, vertical density profile and
distribution of cosine of inclinations of particles from each galactic
component at times $0.5$, $1.0$, $2.0$ and $3.0$ $Gyr/h$.

The average surface mass distributions shows an interesting trend to
form an exponential distribution like a disk. The surface density
profile of gas depicts a distribution clearly exponential between
$\sim 13-90\:kpc/h$. Dark matter, stellar disk and bulge components
exhibit a tendency to form an exponential distribution with peaks in
regions where the rosette has maximum of density. For new stars the
trend is not so strong yet, but particles of this component always get
the behaviour of those from stellar disk component.

Additionally, comparing the shape of the vertical density profiles
with those shown in \autoref{fig:AM_relaxations} reminds very much the
behaviour of a structure with isothermal vertical support. We see that
the different components of the material in the polar structure, all
of them evolve in time getting a vertical distribution with a shape
typical of a thick disk \citep{Binney2008}. This is an indicator of
gravitational support that may suggests a long standing structure. The
structure formed with particles of dark matter acquire this
gravitational support very quickly at $0.5\:Gyr/h$, as this
component has the largest contribution to the mass in the stream this
could guarantee the gravitational support of all the
structure. Particles from the other components get the form of the
vertical distribution after $2\:Gyr/h$.

The gravitational support is generated by the formation of the planar
structure with rotational support that is observed in the distribution
of cosine of inclinations for particles in the stream. Particles in
the stream, specially of gas, begin with orbital inclinations
distributed in all directions with co-rotating and counter-rotating
orbits relative to the rotation of the MG disk, then by conservation
of angular momentum almost all orbits become co-rotating with cosine
of inclination close to one, i.e., all stream forms a structure
perpendicular to the MG disk. This result leads us to claim that
possibly the system AM 2229-735 will evolve to a polar ring galaxy.

\subsection{Dynamical structure of the polar disk}

As it was already mentioned, Figure
\ref{fig:evolution_stream_HR_better_feedback} shows that the average
density distribution of the material in the polar structure settles in
a disk-like structure that follows an almost exponential profile in
the radial direction and in the vertical direction behaves in a way
that suggest the behaviour of an isothermal system.

The simplest solution of Jeans equation for an axisymetric system in
equilibrium with an isothermal energy distribution is

\begin{equation}
  \lambda(z) = \lambda_0 {\rm sech}^2{(z/z_0)},
  \label{eq:vertprofile}
\end{equation}

where $\lambda_0$ is a constant and $z_0$ is the vertical scale length
of the disk and is related to the vertical velocity dispersion of
particles in the disk.  A solution to the problem shows that in
general $\sigma_z(R)$ is not a constant function and depends on the
distance from the centre of the galaxy through the radial surface
density according to

\begin{equation}
  \sigma^2_z(R) \propto \pi G z_0 \Sigma(R),
  \label{eq:veldispscale}
\end{equation}

then in principle, for an isothermal sheet, $\sigma_z(R)$ should
follow an exponential law with a scale length that is twice that of the
surface density profile.

 \begin{figure*}
   \centering
   \includegraphics[scale=0.23,angle=0]{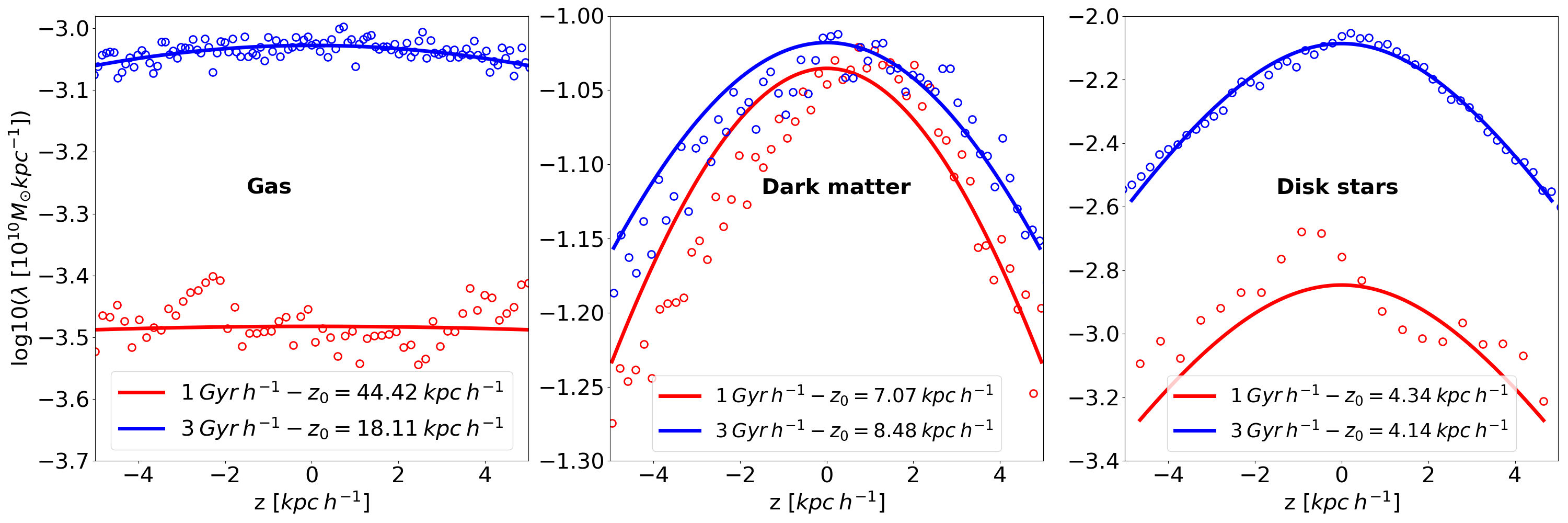}
   \caption{Vertical mass distribution for gas, dark matter and stars
     in polar structure as a function of the height $z$. Lines are fits
     to equation \ref{eq:vertprofile} while the points are the
     data. Red lines correspond to distributions at $1.0\:Gyr/h$ and
     blue lines to $3.0\:Gyr/h$.}
   \label{fig:Vdensfits}
 \end{figure*}

Figure \ref{fig:Vdensfits} shows fits to equation \ref{eq:vertprofile}
of the vertical density profile of the different mass components in
the polar structure. The figure shows the fit for these components at
two different simulation times of $1~Gyr/h$ and $3~Gyr/h$. As it can
be seen in the figure, the vertical structure is very well described
by this profile for the major components of the polar structure, and
as it can be observed in this figure and in figure
\ref{fig:evolution_stream_HR_better_feedback}, the structure tends to
a particular distribution.

In order to study the dynamical conditions of the material in the
stream we study the mean velocity field of the particle
distribution. If we assume that the system is axisymetric, the mean
radial and vertical velocities should be around zero if the system is
in steady state. The azimuthal velocity should be larger than zero
indicating rotational support.

 \begin{figure}
   \centering
   \includegraphics[scale=0.11]{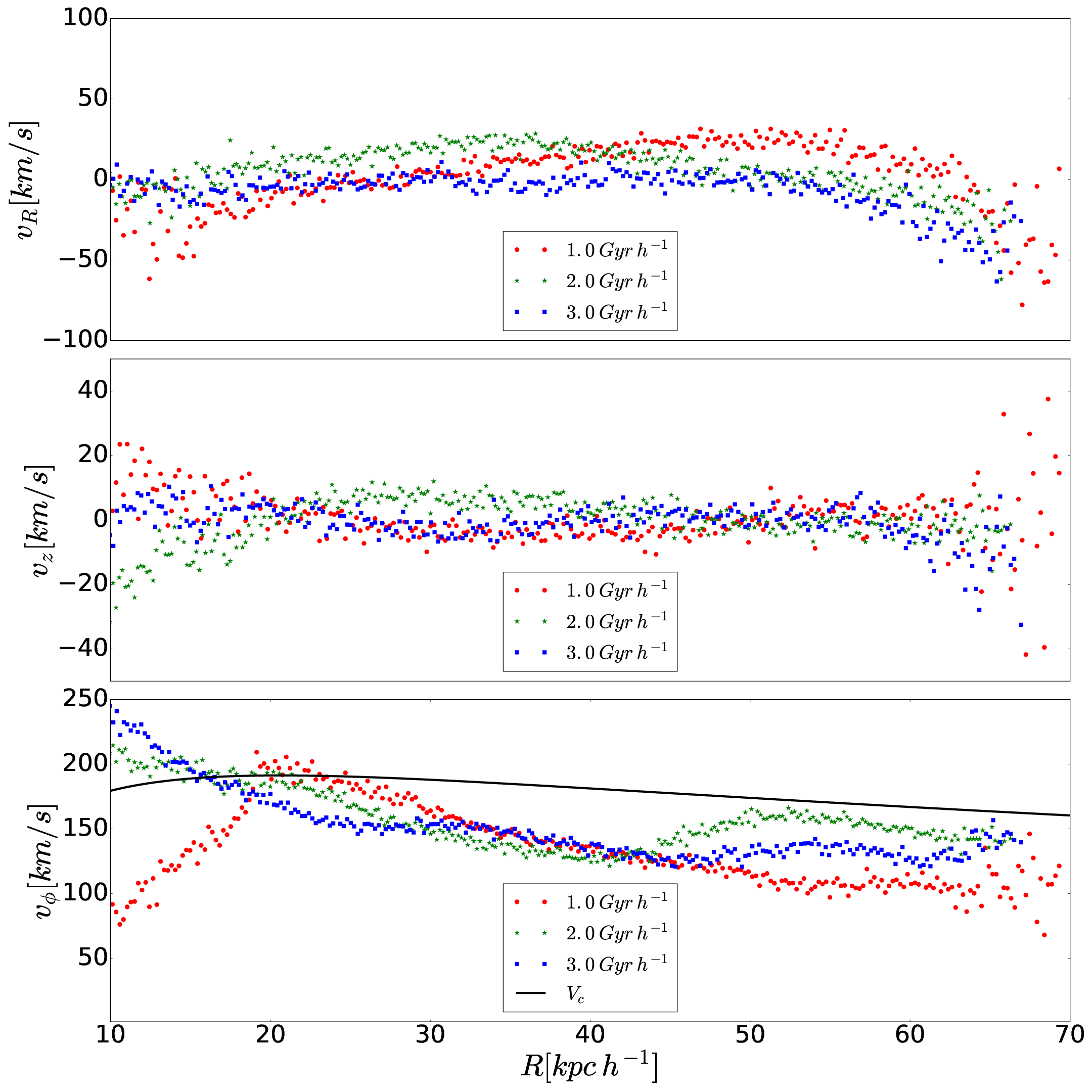}
   \caption{Mean velocity in $R$ (top), $z$ (middle) and $\phi$
     (bottom) as a function of the distance to polar structure centre
     for $1.0$ (red points), $2.0$ (green stars) and $3.0$ (blue
     squares) $Gyr/h$. In bottom panel appear the circular velocity of
     MG dark matter halo with a black line.}
   \label{fig:meanVfield}
 \end{figure}

Figure \ref{fig:meanVfield} shows the mean radial, vertical and
azimuthal velocity $\bar{v}_R$, $\bar{v}_z$ $\bar{v}_{\phi}$ of dark
matter particles in the stream as a function of the distance from the
centre of the disk structure. Similar plots can be obtained for gas
and stars. As it can be seen, the mean vertical and radial velocities
decrease in time to zero at $3~Gyr/h$. Not the same behaviour is
observed for the azimuthal mean velocity that clearly follows a
keplerian-like curve. The solid line shows the estimated circular
velocity of the dark matter halo of the host galaxy in the same radial
regime, showing that the particles in the polar structure tend to move
under the influence of the potential of the host dark matter
halo. That $\bar{v}_R$ and $\bar{v}_z$ tend to zero at latter times
suggest that the structure is going towards a configuration of
equilibrium, and the observed behaviour on $\bar{v}_{\phi}$ indicates
that the equilibrium configuration is supported (or partially
supported) by rotation.

 \begin{figure}
   \centering
   \includegraphics[scale=0.16]{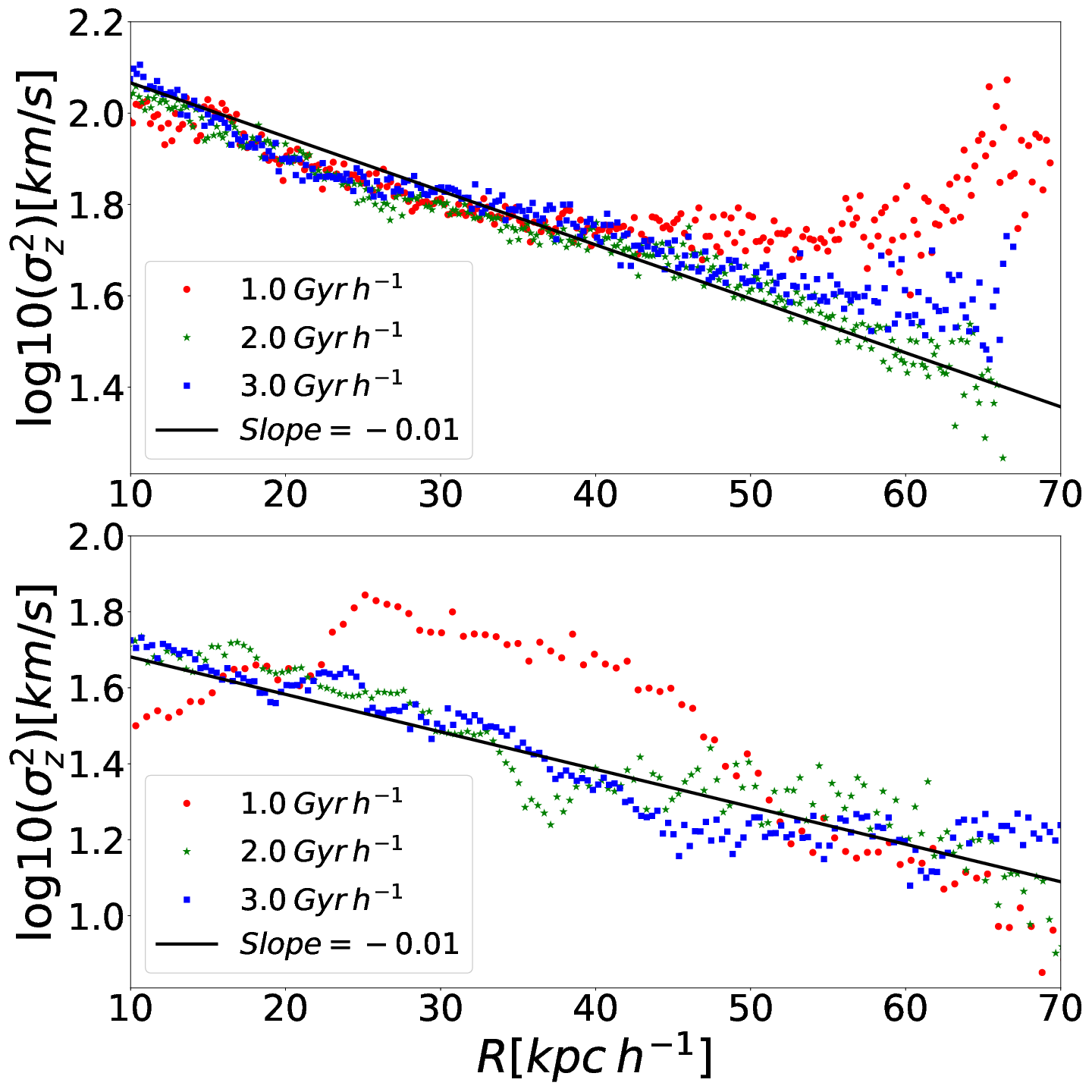}
   \caption{$z$-velocities dispersion vs. the distance to polar
     structure centre for dark matter particles (top panel) and stars
     particles at times of $1.0$ (red points), $2.0$ (green stars) and
     $3.0$ (blue squares) $Gyr/h$. The black lines are fits to straight
     lines.}
   \label{fig:SigmaVz}
 \end{figure}

Figure \ref{fig:SigmaVz} shows the vertical velocity dispersion as a
function of distance from the centre of the polar structure for
different times. As it can be seen in the figure, the larger the time,
the closer the relation to the exponential profile. In agreement with
the previous figures, this shows that the structure is looking for
equilibrium. That $\sigma_z(r)$ follows this behaviour, according to
\autoref{eq:veldispscale}, suggest that the polar structure approaches a
rotational support that at some degree can be approximated with an
isothermal axisymetric system.

Since the HR simulation was ran just until $3~Gyr/h$ of evolution of
the system, we can not see how things evolve in a larger timescale at
this resolution. However LR simulation was ran for up to $10~Gyr/h$, and
even with the low resolution simulation, the system still shows the
same behaviour in such a large period of time.  This result suggest
three interesting implications. First, that polar ring structures
such as those observed in NGC 4650 tend to find equilibrium conditions
that are similar to those of the disk galaxies.  Second, this scenario
and the result of these simulations, in general, are suggesting the
way stellar disks are formed in the hierarchical scenario. Third, and
more important, is that we see in our simulations that there is a major
contribution of dark matter in the mass distribution of the polar
structure. Although it is out of the scope of this work to determine
under which conditions this may be observable, being able to measure
precisely the motion of stars and gas in these structures would
provide evidence of the presence of dark matter in the structure,
turning in a new kind of experimental scenario for the detection of
dark matter in galaxies.

\section{Conclusions}
\label{sec:conclusions}

In this work we have used observational constraints and numerical
simulations to study a possible scenario for the evolution of the real
system AM 2229-735 towards the formation of a polar ring galaxy.

We used the kinematic information of the galaxies in the system AM
2229-735 to apply a method to find a good orbit to simulate their
interaction (\autoref{subsec:orbits}). This method can be used for any
observed minor merger with enough information to estimate the
interaction kinematics and masses of the galaxies in the system.

With this orbit we ran a high resolution simulation and studied the
formation and evolution of a polar structure. We reproduced the main
observed morphological features of system as the arm in main galaxy,
arms in satellite galaxy and the material bridge connecting them,
these features were obtained after the second pass by periastron of SG
at $\sim 1.32$ $Gyr/h$. The tidal stream formed during
interaction contains gas, dark matter, old and young stars coming
mainly from SG by tidal stripping, while the satellite orbits without
being swallowed by the host. This stripped material inherits the
orbital angular momentum from SG and a rosette-like on a plane almost
perpendicular to MG disk is produced forming a galaxy with a polar
structure, a polar ring galaxy.

In order to check for the quality of our results we performed a
careful resolution study running simulations at three different
resolution levels (see \autoref{sec:resolution}). We found that the
global properties of the interaction as the system observation
features and the formation of the polar structure are achieved in the
three simulations. We found that the results from the simulations tend
systematically to a fixed result with increased resolution, as it is
the case of the structure of the orbit (see
\autoref{fig:AM_orbit}). The same trends were observed for other
quantities such as mass accretion and dynamical conditions in the
polar structure. Therefore we are confident that our conclusions are
robust against the effects of numerical discreteness.

In this work we have found that, after modelling the evolution of a
realistic system, it is possible to form a polar disk structure as the
result of a minor merger of galaxies. However, we go one step further
and use the results of our simulations to study the dynamical
conditions of such a structure. In doing so we found evidence that
suggest that these polar disk structures can be approximated as
isothermal axisymetric systems, very much like a classical stellar
disks. We found that in the time, the structure looks for equilibrium
acquiring a vertical structure very much reminiscent of a stellar disk.
This result has different interesting implications. First, knowing the
dynamical nature of the polar structure is itself an interesting
result since it provides information about the physical conditions
under which the material in the structure is evolving. Second, this
process of formation of polar ring structures can be used to
understand the origin of stellar exponential disks in the universe, a
problem that is still yet to be solved. And third, it is interesting
to note that in our simulations there is a major contribution of dark
matter in the polar ring structure. The results presented here suggest
that observations of stellar kinematics of polar ring galaxies may
serve as an scenario to search for dark matter in this kind of
objects, providing a different observational experiment for the search
of dark matter in galaxies.

Finally, our simulations have included feedback, star formation,
etc. We have incorporated these effects in our simulations with the
aim to provide realism to the simulations, but trying to keep the
scope of the work close to the subject of the formation of the polar
structure and its properties, we have not focused our attention on the
subject of the dynamics of gas and star formation in the merger
remnant. In a work under preparation (Quiroga et.al. in prep) we study
the effects of feedback in the formation of stars in the merger
remnant and the polar structure. Despite not considering it in this
work, we are confident that the effects of the baryon physics should
not affect our conclusions since most of the gross dynamics is
dominated by collisionless particles.

\section*{Acknowledgements}

This research work was supported by COLCIENCIAS (doctorados
nacionales, convocatoria 617 de 2013) and the research project
111571250082 (convocatoria 715-2015). N.I.L acknowledges financial
support of the Project IDEXLYON at the University of Lyon under the
Investments for the Future Program (ANR- 16-IDEX-0005). I.R. thanks 
the Brazilian agency CNPq (Project 311920/2015-2). Additionally,
simualtions performed in this work were run in the computer facilities
of GFIF in the Instituto de F\'\i sica, Universidad de Antioquia,
Hipercubo at IP\&D-UNIVAP (FINEP 01.10.0661-00, FAPESP 2011/13250-0 and 
FAPESP 2013/17247-9), Leibniz-Institut F\"ur Astrophysik Potsdam. The
authors gratefully acknowledge the Gauss Centre for Supercomputing
e.V. (www.gauss-centre.eu) for funding this project by providing
computing time through the John von Neumann Institute for Computing
(NIC) on the GCS Supercomputer JURECA at J\"ulich Supercomputing
Centre (JSC).




\bibliographystyle{mnras}
\bibliography{references} 




\appendix

\section{Study of resolution}\label{sec:resolution}

\noindent All results presented in the previous section were obtained
using the high resolution simulation (HR). In this section we study
the implications of doing the same simulation with the same initial
conditions and the same baryonic physics but using lower
resolutions. Remember that these resolutions were defined in Section
\ref{sec:simulations} where HR, MR and LR are tags for simulations of
high, medium and low resolutions. Thus, this experiment shows the
effects of the discretization on the evolution of the system and its
impact in our conclusions.

In left panel of Figure \ref{fig:AM_orbit} we show the
distance between SG and MG as a function of time for the three
different resolutions. Although there are some differences in
trajectories after $1.5$ $Gyr\:h^{-1}$ orbits for HR (red line) and MR
(blue line) resolutions have a very similar behaviour. On the other
hand, although the differences are still small, LR (black line) has a
slightly different behaviour. This produces that the match points are
not fully coincident revealing that the time where simulation
\textit{reproduces} the observation is not the same, however the
differences a very small. The differences in time to math point is of
$0.01\:Gyr$ with LR and $0.02\:Gyr$ with LR, and the differences in
distance are of $0.90\:Kpc$ with MR and $2.45\:kpc$ with LR.

These results are a consequence of the discrete representation of the
system and how the different approximations affect the dynamics of the
system with different number of particles. Then in each resolution the
satellite undergoes unequal gravitational forces and torques that lead
to slightly different trajectories. Obviously, while the number of
particles is increased a better approximation to the continuous is
realised and the discrepancy in orbits decrease like it happens with
HR and LR resolutions.

 \begin{figure}
   \centering
   \includegraphics[scale=0.35,angle=270]{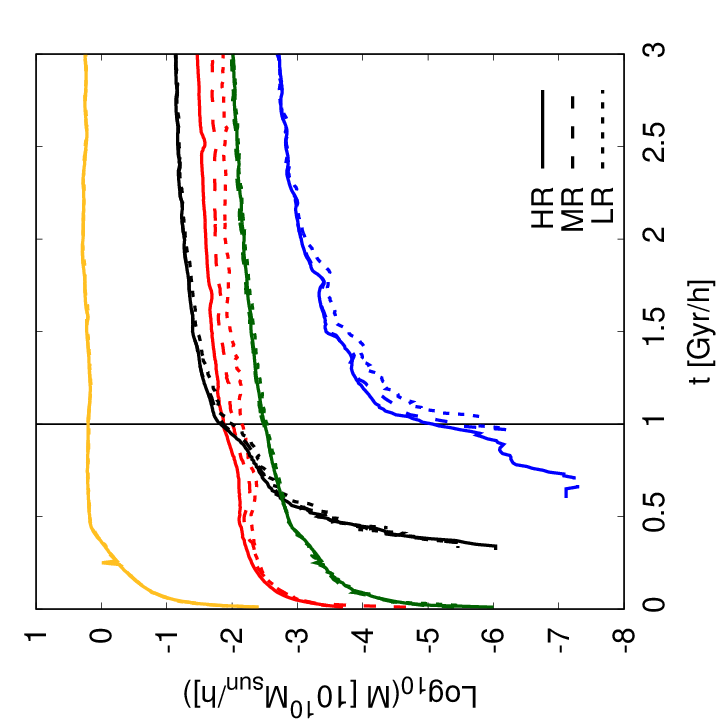}
   \caption{Masses measured for stream for the three resolutions as a
     function of time.  Short dashed lines for LR, dashed lines for MR
     and solid lines for HR. The colours for the mass of each component
     are: red for gas, yellow for dark matter, black for disk stars,
     green for bulge and blue for young stars.}
 \label{fig:mass_resolution}
 \end{figure}

In the same way processes like tidal and ram pressure
stripping and the effects of feedback in the gas particles may be
affected by discretisation. In Figure \ref{fig:mass_resolution} we
show the stream masses measured for each component for the three
resolutions (short dashed lines for LR, dashed lines for MR and solid
lines for HR).

As almost the mass of the polar structure is in dark matter, the first
thing we can see is that the total mass in the polar structure is the
same for all three simulations, independent on the resolution. We see
also that all collisionless components (stellar disk, bulge and DM
halo) have more or less the same mass deposition rate in to the polar
structure. A slight difference is observed for the particles in the
disk in the LR simulation, and it should be due to the effects of
different angular momentum transfer at that low resolution
simulation. Differences are observed for the gas and for the young
stars.

For gas, that undergoes more rich physics via sub-grid model for ISM,
the discretisation is more important than other components producing a
larger discrepancy between its mass curves giving more gas mass to the
stream when the resolution is increased (HR). Since the mass per
particle is lower in HR than in LR then its inertia to external
perturbations is smaller. Also, the resolution affects directly the
star formation and feedback processes such that the amount of cold and
hot gas and young stars available to be part of the stream changes for
each resolution.

The abundance of young stars is slightly different. First, notice that
HR starts showing young stars at earlier times than MR and LR
simulations. This is naturally due to the increased resolution that
allows to track movement of smaller parcels of mass from the galaxies
to the streams and polar structure. Afterwards HR and MR simulations
show more or less the same behaviour, and although at the beginning LR
shows a different rate of mass transfer in to the polar structure, the
final total mass is the same for all simulations.

Despite small differences in orbits and mass accretion of polar
structure, their global features as form and dimensions are similar
implying that orbital properties of particles forming the polar ring
are the same. Obviously, the models subgrid controlling the gas
physics are resolution depending then the physical properties and the
star formation process do not have the same behaviour in simulations.

We argue that we have achieved convergence in our simulations and that
resolution is not affecting our conclusions relative to the formation
and structure of the polar ring forming during the simulations of the
merger of AM 2229-735.


\bsp	
\label{lastpage}
\end{document}